%% file: main.tex
\documentclass[a4paper,fleqn]{cas-dc}

\usepackage[switch,mathlines]{lineno}
\let\oldequation\equation
\let\oldendequation\endequation

\renewenvironment{equation}
  {\linenomathNonumbers\oldequation}
  {\oldendequation\endlinenomath}

\usepackage[numbers]{natbib}

\usepackage{lipsum}

\usepackage{wasysym}

\usepackage{tabularx}
\usepackage{graphicx}
\usepackage{booktabs}
\usepackage{longtable}
\usepackage{caption}
\usepackage{float}
\usepackage{siunitx}
\DeclareSIUnit \decibelA {dB(A)}
\newif\ifarxiv
\arxivtrue

\usepackage{cleveref}

\usepackage{placeins}
\usepackage{flafter}
\usepackage{float}

\usepackage{etoolbox}
\usepackage{xspace}

\newif\ifuseendfloat
\ifuseendfloat
    \usepackage[nolists, nomarkers]{endfloat}
    
\fi

\input{commands/cas-abbrev-fix}
\input{commands/citeabbrev}

\usepackage{pdflscape}
\usepackage{rotating}
\usepackage{longtable}

\usepackage{tikz}
\usetikzlibrary{calc}

\def\SquareSLIDER#1#2{
\tikz[baseline=-0.1cm]{
 \coordinate (start) at (0,-0.1cm);
 \coordinate (end) at (#1,0.1cm);
 \coordinate (mark) at ($(start|-0,0)!#2!(end|-0,0)$);
 \fill[rounded corners=0.1cm, draw=gray, bottom color=lightgray, top color=black, middle color=lightgray] (start) rectangle (end);
 \shade[draw=darkgray, rounded corners=0.2mm, ball color=gray!20!lightgray] (mark) +(-.15,-.15) rectangle +(.15, .15) ;
}
}

\usepackage{xcolor}
\definecolor{few_pal_1}{RGB}{93,165,218}
\definecolor{few_pal_2}{RGB}{250,164,58}
\definecolor{few_pal_3}{RGB}{96,189,104}
\definecolor{few_pal_4}{RGB}{241,124,176}

\definecolor{brbg_pal_1}{RGB}{140,81,10}
\definecolor{brbg_pal_2}{RGB}{191,129,45}
\definecolor{brbg_pal_3}{RGB}{223,194,125}
\definecolor{brbg_pal_8}{RGB}{53,151,143}
\definecolor{brbg_pal_9}{RGB}{1,102,94}

\usepackage{hyperref}

\input{citealias}

\begin{document}\sloppy
\let\WriteBookmarks\relax
\def\floatpagepagefraction{1}

\input{commands/malay}

\input{sections/00front}

\ifarxiv\else\linenumbers\fi

\input{sections/01intro}

\input{sections/02method}

\input{sections/03listening}

\input{sections/04circumplex}

\input{sections/05discussion}

\input{sections/06conclusion}

\input{sections/99endmatter}

\input{sections/98appendix}

\end{document}

%% file: commands/cas-abbrev-fix.tex
\ExplSyntaxOn
\RenewDocumentCommand \eadauthor {} 
    { 
      \seq_map_inline:Nn \l_stm_au_seq 
        { 
            \regex_extract_once:nnNTF {(\w)\w*-(\w)} { ##1 } \l_stm_au_fn_seq
            { 
                \seq_pop_left:NN \l_stm_au_fn_seq \temp_var
                \seq_use:Nn \l_stm_au_fn_seq { .- }
                { . } 
            }
            { 
                \regex_match:nnTF { \. } { ##1 } 
                { ##1 }
                { \tl_head:n {##1}. }
            }
      }{ ~\l_stm_au_sn_seq }
    }
\ExplSyntaxOff

%% file: commands/citeabbrev.tex
\newif\ifabbreviation
\pretocmd{\thebibliography}{\abbreviationfalse}{}{}
\AtBeginDocument{\abbreviationtrue}
\DeclareRobustCommand\acroauthor[2]{%
  \ifabbreviation
    \ifcsname acroused@#2\endcsname
      #2%
    \else
      #1%
      ~[\mbox{#2}]
      \expandafter\gdef\csname acroused@#2\endcsname{}%
    \fi
  \else
    \ifcsname bibacroused@#2\endcsname
        \mbox{#2}%
    \else
        \mbox{#1}~(\mbox{#2})%
        \expandafter\gdef\csname bibacroused@#2\endcsname{}%
    \fi
  \fi
}

%% file: citealias.tex
\defcitealias{InternationalOrganizationforStandardization2014ISOFramework}{ISO, 2014} 

\defcitealias{InternationalOrganizationforStandardization2018ISO/TSRequirements}{ISO, 2018} 

\defcitealias{InternationalOrganizationforStandardization2019ISO/TSAnalysis}{ISO, 2019}

\defcitealias{InternationalOrganizationforStandardization2007ISOLanguage}{ISO, 2007}

%% file: commands/malay.tex
\newcommand{\candidate}[1]{\textlangle\textit{#1}\textrangle}
\newcommand{\ccandidate}[2]{\textlangle\textsc{#1}: \textit{#2}\textrangle}

\newcommand{\andcomp}[4]{\ccandidate{#1}{#2}\ and \ccandidate{#3}{#4}}

\newcommand{\ccomp}[4]{\andcomp{#1}{#2}{#3}{#4}}

\newcommand{\pcomp}[2]{\candidate{#1}\ and \candidate{#2}}

\makeatletter
\newcommand{\defcand}[2]{%
  \@namedef{#1}{#2}%
  \@namedef{i#1}{\textit{#2}}%
  \@namedef{c#1}{\candidate{#2}}%
  \@namedef{sc#1}{\textsc{#2}}%
  \@namedef{sg#1}{\ccandidate{sg}{#2}}%
  \@namedef{my#1}{\ccandidate{my}{#2}}%
  \@namedef{insg#1}{\textit{#2} in 
  \textsc{sg}}
  \@namedef{inmy#1}{\textit{#2} in 
  \textsc{my}}
  \@namedef{sc#1}{\textsc{#2}}
  \@namedef{mc#1}{\mathcal{#2}_n}
  \@namedef{cal#1}{\mathcal{#2}}
}
\makeatother

\defcand{sn}{S}
\defcand{s}{S}

\defcand{isopl}{ISOPL}
\defcand{isoev}{ISOEV}

\defcand{bm}{Bahasa Melayu}
\defcand{seba}{Sebutan Baku}

\defcand{ell}{ell} 
\defcand{eng}{eng} 
\defcand{zsm}{zsm} 
\defcand{tha}{tha} 
\defcand{por}{por} 
\defcand{deu}{deu} 
\defcand{spa}{spa} 

\defcand{mym}{my:m}
\defcand{myo}{my:o}
\defcand{sg}{sg}
\defcand{my}{my}
\defcand{sgq}{sq:q}
\defcand{myq}{my:q}

\defcand{e}{eventful}
\defcand{v}{vibrant}
\defcand{p}{pleasant}
\defcand{ca}{calm}
\defcand{u}{uneventful}
\defcand{m}{monotonous}
\defcand{ch}{chaotic}
\defcand{a}{annoying}

\defcand{membi}{membingitkan}
\defcand{menje}{menjengkelkan}

\defcand{meriah}{meriah}

\defcand{menye}{menyenangkan}

\defcand{timer}{tidak meriah}

\defcand{menen}{menenangkan}
\defcand{tenang}{tenang}

\defcand{huru}{huru-hara}
\defcand{kelam}{kelam-kabut}

\defcand{membo}{membosankan}
\defcand{tiber}{tidak berubah oleh itu membosankan}
\defcand{kurang}{kurang kepelbagaian oleh itu membosankan}

\defcand{berse}{bersemarak}
\defcand{rancak}{rancak}

\defcand{ncon}{ncon}
\defcand{ortho}{ortho}
\defcand{ibal}{ibal}

%% file: sections/00front.tex
\newcommand*{\papertitle}{Crossing the Linguistic Causeway: Ethnonational Differences on Soundscape Attributes in Bahasa Melayu}
\shorttitle{\papertitle}    
\shortauthors{B. Lam et al.}  

\title[mode=title]{\papertitle}

\input{authoraffil}
\tnotemark[1]
\tnotetext[1]{The research protocols used in this research were approved by the respective institutional review board of Nanyang Technological University (NTU), Singapore [IRB-2021-293] and Universiti Putra Malaysia (UPM), Malaysia [JKEUPM-2019-452].}

\begin{abstract}
Despite being neighbouring countries and sharing the language of Bahasa Melayu (ISO~639-3:~\sczsm), cultural and language education policy differences between Singapore and Malaysia led to differences in the translation of the ``\textit{annoying}'' perceived affective quality (PAQ) attribute from English (ISO~639-3:~\sceng) to \sczsm. This study expands upon the translation of the PAQ attributes from \sceng\ to \sczsm\ in Stage 1 of the Soundscapes Attributes Translation Project (SATP) initiative, and presents the findings of Stage 2 listening tests that investigated ethnonational differences in the translated \sczsm\ PAQ attributes and explored their circumplexity. A cross-cultural listening test was conducted with 100 \sczsm\ speakers from Malaysia and Singapore using the common SATP protocol. The analysis revealed that Malaysian participants from non-native ethnicities (\scmyo) showed PAQ perceptions more similar to Singapore (\scsg) participants than native ethnic Malays (\scmym) in Malaysia. Differences between Singapore and Malaysian groups were primarily observed in stimuli related to water features, reflecting cultural and geographical variations. Besides variations in water source-dominant stimuli perception, disparities between \scmym\ and \scsg\ could be mainly attributed to \iv\ scores. The findings also suggest that the adoption of region-specific translations, such as \imembi\ in Singapore and \imenje\ in Malaysia, adequately addressed differences in the \ia\ attribute, since significant differences were observed in one or fewer stimuli across ethnonational groups. The circumplexity analysis indicated that the quasi-circumplex model better fit the data compared to the assumed equal angle quasi-circumplex model in ISO/TS~12913-3, although deviations were observed possibly due to respondents' unfamiliarity with the United Kingdom-centric context of the stimulus dataset. Furthermore, the alignment between Stage 2 listening tests and quantitative evaluation of attributes in Stage 1 revealed biases in the \ie--\iu\ dimension across ethnonational groups. This study provides insights into the perception of PAQ attributes in cross-cultural and cross-national contexts, facilitating the culturally appropriate adoption of translated PAQ attributes in soundscape evaluation.
\end{abstract}

\ifarxiv\else
\begin{highlights}
\item Across all stimuli, \scmyo\ participants were closer to \scsg\ than \scmym\ in PAQ perceptions
\item Country-specific translations of \ia\ effectively addressed cultural distinctions
\item Water-related stimuli influenced PAQ attribute differences among ethnonational groups
\item Limited stimulus diversity may explain non-fit to equal angle quasi-circumplex model
\item Circumplex skew in \ie\ for \scmym\ and \iv\ for all groups deviated from Stage 1
\end{highlights}
\fi

\maketitle

%% file: authoraffil.tex
\author[eee]{Bhan Lam}[orcid=0000-0001-5193-6560,
degree=Ph.D.]
\ead{blam002@e.ntu.edu.sg}
\corref{c}\cortext[c]{Corresponding authors}
\credit{Conceptualization, Methodology, Software, Validation, Formal analysis, Investigation, Project administration, Data Curation, Writing - Original Draft, Writing - Review \& Editing, Visualization, Supervision}

\author[upm]{Julia Chieng}[orcid=0000-0003-1407-3238, degree=Ph.D.]
\ead{chiengjulia@upm.edu.my}
\corref{c}
\credit{Conceptualization, Methodology, Validation, Investigation, Resources, Writing - Review \& Editing, Project administration}

\author[eee]{Kenneth Ooi}[orcid=0000-0001-5629-6275]
\ead{wooi002@e.ntu.edu.sg}
\credit{Software, Formal analysis, Resources, Writing - Review \& Editing}

\author[eee]{Zhen-Ting Ong}[orcid=0000-0002-1249-4760]
\ead{ztong@ntu.edu.sg}
\credit{Resources, Project administration}

\author[gt,eee]{Karn~N. Watcharasupat}[orcid=0000-0002-3878-5048]
\ead{kwatcharasupat@gatech.edu}
\credit{Methodology, Writing - Review \& Editing}

\author[cnu]{Joo Young Hong}[orcid=0000-0002-0109-5975, degree=Ph.D.]
\ead{jyhong@cnu.ac.kr}
\credit{Resources, Funding acquisition, Writing - Review \& Editing}

\author[eee]{Woon-Seng Gan}[orcid=0000-0002-7143-1823, degree=Ph.D.]
\ead{ewsgan@ntu.edu.sg}
\credit{Resources, Writing - Review \& Editing, Supervision, Funding acquisition}

\affiliation[eee]{
    organization={
        School of Electrical and Electronic Engineering, 
        Nanyang Technological University%
    },
    addressline={50 Nanyang Ave, S2-B4a-03}, 
    postcode={639798}, 
    country={Singapore}
}

\affiliation[upm]{
    organization={
        Department of Music,
        Faculty of Human Ecology,
        Universiti Putra Malaysia%
    },
    addressline={43400 UPM Serdang}, 
    state={Selangor Darul Ehsan},
    country={Malaysia}
}

\affiliation[gt]{
    organization={
        Center for Music Technology,
        Georgia Institute of Technology%
    },
    addressline={J. Allen Couch Building, 840 McMillan St NW}, 
    city={Atlanta},
    postcode={30332}, 
    state={GA},
    country={USA}
}

\affiliation[cnu]{
    organization={
        Department of Architectural Engineering,
        Chungnam National University%
    },
    addressline={34134}, 
    city={Daejeon},
    country={Republic of Korea}
}

%% file: sections/01intro.tex
\section{Introduction}\label{sec:intro}

The Soundscape Attributes Translation Project (SATP) is a global initiative to develop methodological translations to eight perceived affective quality (PAQ) attributes that describe the overall soundscape quality in the ISO 12913 series of standards \cite{Aletta2020}. In Stage 1 of the SATP initiative, a binational (Malaysia and Singapore) expert-led approach augmented by a quantitative strategy was employed to select a provisional set of translations for \ibm\ (ISO~639-3:~\sczsm) \cite{Lam2022MalayP1}. This article describes Stage 2 of the SATP initiative, whereby the provisional translations are used to evaluate a standardized but acoustically diverse set of audio stimuli. Due to the multicultural and multiethnic composition of the \sczsm\ speakers in Malaysia and evolving linguistic and cultural ties between Singapore and Malaysia, the evolution of \ibm\ in both countries is provided for context in \Cref{sec:evolutionBM}, which could later account for any ethnonational differences. 

\subsection{Evolution of \ibm\ in Singapore and Malaysia} \label{sec:evolutionBM}

Malaysia (\scmy) and Singapore (\scsg) are multilingual and multiethnic South-East Asian maritime neighbouring countries where Standard Malay (\ibm) (ISO~639-3:~\sczsm) is a national and official language. Singapore is a densely populated island nation with a land area of \SI{734.3}{\kilo\meter\squared} and a population of 5.64 million (2022, inclusive of ``non-residents''). Malaysia is Singapore's closest maritime neighbour, with a much larger land area of \SI{330524}{\kilo\meter\squared} and a population of 33 million (2022). The two countries are connected by two bridges, the Johor--Singapore causeway, one of the busiest border crossings in the world, and the Malaysia--Singapore Second Link.

Ethnically, Singapore's 4.07 million resident population is majority Chinese (\SI{74.18}{\percent}), followed by Malay (\SI{13.61}{\percent}), Indian (\SI{8.99}{\percent}) and Others (\SI{3.29}{\percent}) \cite{SingaporeDepartmentofStatistics2022}, whereas Malaysia's 30.4 million citizens are majority Malay (\SI{57.89}{\percent}), followed by Chinese (\SI{22.70}{\percent}), Other Indigenous (\SI{12.17}{\percent}), Indian (\SI{6.58}{\percent}), and Others (\SI{0.66}{\percent}) \cite{DepartmentofStatisticsMalaysia2023}.

The commonality of the \sczsm\ variety in Singapore and Malaysia dates back to the Johor-Riau-Lingga Sultanate (17th to 19th century), where the Johor-Riau dialect was the \textit{lingua franca}. Eventually, the Johor-Riau dialect formed the basis for the development of the written and naturalised spoken form (\textit{Sebutan Johor-Riau}) of Standard Malay (\sczsm) in Malaysia and Singapore throughout the 19th to 20th century \cite{Bakar2021}. 

In an attempt to unify the Malay-speaking communities in the Malay Peninsula and Archipelago (e.g. Malaysia, Singapore, Brunei, Indonesia), an artificially created system of pronunciation (\iseba) was proposed in the 1980s, which sounds similar to Bahasa Indonesia (ISO~639-3:~\textsc{ind}) but distinct from \textit{Sebutan Johor-Riau} (\sczsm) \cite{Bakar2021}. Even though \iseba\ was fully implemented in Malaysia and Singapore in 1988 and 1993, respectively, Malaysia eventually reinstated \textit{Sebutan Johor-Riau} in 2000 due to potential loss of national identity. Despite a similar push-back and falling through of the standardisation, the \iseba\ policy was not retracted in Singapore, giving rise to a young population ($<40$ years old) of native speakers of the artificial \iseba\ dialect of \sczsm.

Due to the bilingual education policy in Singapore \citep{Lam2022MalayP1, Alfred1987}, where English is the ``first'' language, majority of the population gravitates towards a mother tongue (``second'' language) that reflects their ethnicity, e.g., Mandarin Chinese and Standard Malay for ethnic Chinese and Malays, respectively. Hence, although \sczsm\ is one of the four official languages in Singapore (including English, Mandarin Chinese, Tamil), \sczsm\ speakers are generally ethnic Malays. This is evidenced in the latest census (2020) in Singapore \cite{DepartmentofStatisticsSingapore2020}, where \sczsm\ is spoken most frequently in ethnic Malay households (\SI{60.7}{\percent}) out of other languages, but only within a small number of Indian (\SI{6.0}{\percent}) and Chinese (\SI{0.2}{\percent}) households. In Malay households, however, English (\SI{17.0}{\percent} in 2010; \SI{39.0}{\percent} in 2020) appears to be replacing \sczsm\ (\SI{82.7}{\percent} in 2010; \SI{60.7}{\percent} in 2020) as the most frequently spoken language at home in the last 10 years. Nevertheless, the Malay ethnic community has maintained a high level of proficiency (\SI{90}{\percent}) in \sczsm, across all age groups \citep{Mathews2020}. 

In Singapore, there is a distinction between the ``national'' language (of which \sczsm\ is the only one) and an ``official'' language (of which \sczsm\ is one of the four stated in the preceding paragraph) \cite{lky1965statement}. As a national language, \sczsm\ is used for symbolic purposes like the national anthem, whereas as an official language, \sczsm\ is used in daily and governmental functions in conjunction with the other official languages of Singapore.

On the other hand, \textit{Bahasa Melayu} (\sczsm; also termed \textit{Bahasa Malaysia} \cite{HongSim2019}), is the sole national and official language in Malaysia as legislated in the National Language Act 1963/67 (revised 1971). Within the act, English is permitted in legislative and parliamentary settings, which cements the status of English as the \textit{de facto} second language. Despite the nationalisation of \sczsm\ in Malaysia, the proficiency of \sczsm\ is much higher in ethnic Malays (\SI{84}{\percent}) than among Chinese (\SI{63}{\percent}) and Indians (\SI{57}{\percent}) \cite{MinistryofEducationMalaysia2013}. This discrepancy could be attributed to the medium of instruction in the national-type primary schools. Whereas about \SI{76.23}{\percent} of all national primary schools use \sczsm\ as the medium of instruction, the remainder of the schools use Mandarin Chinese (\SI{16.92}{\percent}) and Tamil (\SI{6.85}{\percent}) as the medium of instruction \cite{EducationalPolicyPlanningandResearchDivision2022}. Moreover, Mandarin Chinese and Chinese dialects are more prevalent in daily conversations than \sczsm\ among ethnic Chinese.

The cross-national divergence of \sczsm\ between Malaysia and Singapore, especially for the younger Singaporeans, and inter-ethnic differences in \sczsm\ proficiency between ethnic Malays and other races in Malaysia may affect the interpretation of certain emotive and affective words, including those used to describe sound. For instance, in Stage 1, cross-national differences were uncovered for the translation of \ia\ from English (ISO~639-3:~\sceng) to \sczsm\ \cite{Lam2022MalayP1}. 

\subsection{Circumplexity of translated perceived affective quality attributes} \label{sec:intro-circ}

In the ISO 12913 series of standards, the affective quality of a soundscape is quantified by an 8-attribute PAQ scale based on the Swedish Soundscape Quality Protocol (SSQP) \cite{Axelsson2010,Axelsson2015HowQuality}. While not explicitly stated in the standards, the PAQ scale is described in the referenced literature as a circumplex model. This is further illustrated by the circular representation in ISO 12913-3 (Fig. A.1 in \cite{iso12913-3}), along with the formulae for the ``Pleasantness'' and ``Eventfulness'' scores that rely on a fixed \SI{45}{\degree} separation between adjacent attributes on the circular model (Eqs. A.1 and A.2 in \cite{iso12913-3}). The validity of these scores, respectively referred to as \iisopl\ and \iisoev\ here, depends on the ``equal angles'' assumption, which is often not verified in practice. Circular arrangement of attributes with equal angular spacing but without being equidistant from the center is usually referred to as a \textit{quasi-circumplex model with equal angles} \cite{Tracey2000}, discussed further in \Cref{sec:circquasi}.

When translating to other languages, PAQ attributes corresponding to the ``main'' axes (i.e. \ia--\ip, \ie--\iu) have been found to be consistent \cite{Nagahata2020,Jeon2018AExperiments}. In cross-lingual studies, however, ``derived'' axis attributes (i.e. \iv--\im, \ica--\ich) appear to deviate from their axial positions on the circumplex \cite{Moshona2023}. Although variations in the derived axis attributes were observed in earlier cross-lingual studies, such as semantic similarities between \im--\iu\ and \textit{exciting}--\ie\ in Korean, these differences could be attributed to imperfect translations from English to the target languages \cite{Jeon2018AExperiments,Moshona2023}. 

Recent studies emerging from the SATP initiative, employing more reliable translation methodologies, have revealed notable differences, especially in the derived axis attributes. In Greek (ISO~639-3:~\scell) translations of the PAQs, \iv\ in \scell\ leaned towards \ie\ in \sceng\ on the circumplex; and a greater deviation in \ich\ was observed between \scell\ and \sceng\ participants on the circumplex \cite{Papadakis2022}. For translations in Portuguese (ISO~639-3:~\scpor), significant differences were found between \scpor\ and \sceng\ for participants in Portugal (PT) and United Kingdom (UK) concerning \iv, \ip\ and \iu; and between \scpor\ in Brazil (BR) and \sceng\ in the UK across \iu\ and \ich\ \cite{MonteiroAntunes2023}. Cross-cultural differences were also observed between \scpor\ participants in PT and BR across all derived axis attributes \cite{MonteiroAntunes2023}; and between \sczsm\ participants in Singapore and Malaysia solely in \ia\ \cite{Lam2022MalayP1}. Ignoring such linguistic and cultural deviations may lead to misinterpretations when computation of \iisopl\ and \iisoev\ attributes.

Nevertheless, attempts have been made to maximize the circumplexity fit of the PAQ attributes during SATP Stage 1 translations in German (ISO~639-3:~\scdeu) \cite{Moshona2023}, Thai (ISO~639-3:~\sctha) \cite{Watcharasupat2022a}, and Bahasa Melayu (ISO~639-3:~\sczsm) \cite{Lam2022MalayP1}. Furthermore, an analysis of circumplexity was conducted using principal components analysis in \scell, while the structural summary method was employed in Spanish (ISO~639-3:~\scspa) \cite{Vida2023}.

\subsection{Research questions} \label{sec:rq}

Building from the quantitative evaluation of the \sczsm\ translations in Stage 1 of the SATP across native \sczsm\ speakers in Singapore (\textsc{sg}) and Malaysia (\scmym) \cite{Lam2022MalayP1}, this work seeks to validate both sets of \sczsm\ translations from Stage 1 through a common listening test protocol, i.e. Stage 2 of the SATP. To determine the applicability of the translated PAQ attributes in \sczsm\ across the entire Malaysian population, the effect of \sczsm\ proficiency in other non-native Malaysian \sczsm\ speakers (\textsc{my:o}) on the soundscape evaluations are also investigated. 

Hence, ethnonational differences in the translated PAQ attributes in \sczsm\ are examined for each PAQ attribute across ethnonational groups and across acoustically diverse soundscapes. Specifically, this study investigates the following research questions:
\begin{enumerate}
  \item To what extent are ethnonational differences in the \sczsm-speaking populations across Malaysia and Singapore influencing soundscape perception in relation to methodologically translated PAQ attributes?
  \item Do the \sczsm\ PAQ attributes exhibit adherence to the implied circumplex structure in the formulation of the \textit{Pleasantness} and \textit{Eventfulness} scores, as outlined in ISO 12913-3:2019?
  \item Do the factor loadings of the Stage 2 listening tests corroborate well with the quantitative method attributes from Stage 1?
\end{enumerate}

%% file: sections/02method.tex

\section{Methods}\label{sec:method}

After the provisional translations of the perceived affective quality (PAQ) attributes to Bahasa Melayu (SATP Stage 1) \cite{Lam2022MalayP1}, listening tests were conducted based on a standardised protocol with a common dataset of audio stimuli (SATP Stage 2) \cite{Aletta2020}.  

\subsection{SATP listening experiment protocol and stimuli}

The SATP Stage 2 listening experiments take place in controlled laboratory settings, with participants using headphones to listen to audio-only stimuli. The translated questionnaire can be administered either electronically or through paper printouts. In the case of electronic surveys, it is recommended to utilise a 100-step slider initialised at 50 to evaluate each PAQ attribute.

To ensure economic accessibility of SATP while maintaining reliability and repeatability, a cost-effective open circuit voltage (OCV) headphone calibration procedure was developed and mandated for the Stage 2 listening experiments \cite{Lam2022ICSV,Lam2022ICA}. The audio stimuli form a shared dataset containing \num{27} excerpts of soundscape recordings, each \num{30}-second in length and encompassing a wide range of sound pressure levels (SPL) from \num{40} to \SI{80}{\decibelA}, as well as PAQs \cite{Oberman2022}. These recordings were conducted in accordance with the protocol outlined in \cite{Mitchell2020}, specifically in London, United Kingdom (UK). An independent characterisation of the sound stimuli was performed by the \scpor\ translation team and documented in \cite{MonteiroAntunes2023}.

\subsection{Study site and administration}

This cross-national study was carried out in Nanyang Technological University (NTU) in Singapore and Universiti Putra Malaysia (UPM) in  Malaysia. Formal ethical approvals were obtained from the NTU Institutional Review Board (IRB-2021-293) and the UPM Ethics Committee For Research Involving Human Subjects (JKEUPM-2019-452). 

The listening experiments in NTU took place in a listening room, while a recording studio was was utilised in UPM. Due to resource constraints, different audio equipment was used for the OCV calibration at each location, as outlined in \Cref{tab:equip}. Nevertheless, the stimuli were calibrated in accordance to the stipulated protocols in both NTU and UPM \cite{Lam2022ICSV}. Additionally, all the listening experiments were conducted using the same MATLAB-based graphical user interface (GUI) \cite{Ooi_GUI_for_SATP}.

\input{table/equipment}
\input{table/demo}

\subsection{Study experiment design}\label{sec:studyexptdesign}

Listening tests were conducted to examine the ethnonational differences in \textsc{zsm} soundscape attribute evaluations between the ethnically homogeneous \textsc{zsm}-speaking population in Singapore, the ethnic Malays in Malaysia, and non-Malays in Malaysia. 

Each \num{30}-second stimulus was presented over headphones and evaluated electronically through a MATLAB-based GUI \cite{Ooi_GUI_for_SATP} on a laptop. The same procedure was repeated for all 27 such stimuli. Participants assessed each stimulus on 8 perceived affective quality (PAQ) attributes (i.e. \textit{eventful}, \textit{vibrant}, \textit{pleasant}, \textit{calm}, \textit{uneventful}, \textit{monotonous}, \textit{annoying}, \textit{chaotic}) in \textsc{zsm} as translated in SATP Stage 1 \cite{Lam2022MalayP1}, and shown in \labelcref{tab:inittransPAQ}. The attributes were evaluated on a 101-point scale [Strongly disagree (0) -- Strongly agree (100)], and were initialised at 50. 

In the GUI, participants were instructed to evaluate the tracks only after listening to the entire track at least once. Moreover, the sliders for each track were programmed to appear only after the entire track had been presented to the participant at least once. Participants were free to repeat the tracks as many times as required to complete the evaluation of the 8 attributes via sliders.

Due to a technical error in the GUI, the majority of the participants in NTU ($n=24$) were presented the stimuli in different random orders, whereas a portion ($n=10$) were presented the stimuli in a fixed random order. On the other hand, all the participants in UPM were presented the stimuli in a fixed random order.

\subsection{Participants}
A total of 100 participants were recruited through convenience and snowball sampling methods for the listening experiments, of which 34 were recruited in Singapore and 66 in Malaysia. The reported demographics of the included study participants are summarised in \Cref{tab:demo}.

 In Singapore (\textsc{sg}), participants were first screened for self-reported spoken fluency in \textsc{zsm} and only invited if they self-reported as fluent (``yes''). After the data collection, one participant was excluded due to hearing loss and another due to a technical error. All remaining 32 participants had normal hearing as screened with an audiometric test (Interacoustics AD629), for all the frequencies tested (mean threshold of hearing < 15 dB at 0.125, 0.5, 1, 2, 3, 4, 6, and 8 kHz). The \textsc{sg} participants had equal gender distribution (Female: 16, Male: 16) and were generally young ($\mu_\text{age, \textsc{sg}}=\SI{25.91}{years}$, $\sigma_\text{age, \textsc{sg}}=7.18$).

Although no audiometric tests were administered for the participants recruited in Malaysia, 3 participants were excluded on the basis of self-reported hearing loss. Of the remaining 63 participants, about half were ethnic Malays (\textsc{my:m}, 31) and the other approximate half were of other ethnicities (\textsc{my:o}, 32). There was approximately equal gender distributions in \textsc{my:m} (Female: 15, Male: 16) and \textsc{my:o} (Female: 16, Male: 16), and both groups were generally young ($\mu_\text{age, \textsc{my:m}}=24.00$, $\sigma_\text{age, \textsc{my:m}}=4.87$; $\mu_\text{age, \textsc{my:o}}=23.09$, $\sigma_\text{age, \textsc{my:o}}=2.64$).

 It is worth noting that the self-reported spoken fluency was a binary assessment in \textsc{sg}, but was rated on a 11-point scale in both \textsc{my:m} ($\mu_\text{spoken, \textsc{my:m}}=9.48$, $\sigma_\text{spoken, \textsc{my:m}}=0.72$) and \textsc{my:o} ($\mu_\text{spoken, \textsc{my:m}}=6.16$, $\sigma_\text{spoken, \textsc{my:m}}=1.53$). On the assumption that a score of at least 7 indicated fluency for \textsc{my} groups, all \textsc{sg} and \textsc{my:m} participants reported fluency in spoken \textsc{zsm}, but only \SI{53.13}{\percent} of the \textsc{my:o} participants reported fluency. To preserve ethnonational differences, spoken fluency was not employed as an exclusion criteria in the \textsc{my:o} group. The self-reported written fluency was similar between \textsc{sg} ($\mu_\text{written, \textsc{sg}}=8.06$, $\sigma_\text{written, \textsc{sg}}=1.58$) and \textsc{my:m} ($\mu_\text{written, \textsc{my:m}}=8.74$, $\sigma_\text{written, \textsc{my:m}}=1.26$), but much lower in \textsc{my:o} ($\mu_\text{written, \textsc{my:o}}=6.38$, $\sigma_\text{written, \textsc{my:o}}=1.56$). 

\subsection{Data analysis}

To assess the potential influence of order effects, a comparison was made between the distributions of the fixed order and random order group using a non-parametric two-sample Kolmogorov-Smirnov (KS) test. To control for multiple comparisons, the \textit{p}-values from the KS test were adjusted using the Benjamin-Hochberg (BH) correction --  a widely accepted method for controlling the false discovery rate. 

Due to unequal sample sizes across the \textsc{sg}, \textsc{my:m} and \textsc{my:o} listening test groups, the non-parametric Kruskal-Wallis test (KWT) \cite{Kruskal1952UseAnalysis} was employed to examine the differences between groups across all 8 PAQ attributes, as well as the \iisopl\ and \iisoev\ attributes. Where differences were found in the KWT at a \SI{5}{\percent} significance level, the pairwise differences were investigated with the posthoc Conover-Iman test (CIT) \cite{Conover1979}, also at a \SI{5}{\percent} significance level.

Upon determining the suitability of factorial analysis with the Kaiser-Meyer-Olkin (KMO) test ($>0.7$) and Barlett's test of sphericity ($p<0.05$), a principal components analysis (PCA) based on the singular value decomposition of data matrix was conducted across the 8 PAQ attributes for each ethnonational group. 

Subsequently, the circumplexity of the 8 attributes projected onto the first two principal components (as was performed to obtain the underlying ISO/TS 12913-3 PAQ model) was examined in terms of circular-order fit with the randomised test of hypothesised order relations (RTHOR) \citep{Hubert1987,Tracey2000}; circulant fit with the structural summary method (SSM) \cite{Zimmermann2017}; and covariance structure modeling approach to circumplexity testing through the Comparative Fit Index (CFI), Root Mean Square Error Approximation (RMSEA), and the Standardised Root Mean Square Residual (SRMSR) \cite{Moshona2023,Grassi2010a}.

All data analyses were conducted with the R programming language \citep{RCoreTeam2021} on a 64-bit ARM environment. The analyses were performed with these specific R packages: KS and BH with \texttt{stats} \cite{RCoreTeam2021}; KWT with \texttt{stats} and \texttt{rstatix} \cite{RCoreTeam2021,Kassambara2021}; CIT with \texttt{conover.test} \cite{AlexisDinno2017}; KMO and Bartlett's test with \texttt{psych} \cite{Revelle2021}; PCA with \texttt{stats} and \texttt{factoextra} \cite{Kassambara2020}; RTHOR with \texttt{RTHORR} \cite{Tracey2020}; SSM with \texttt{circumplex} \cite{Girard2021}; and CFI, RMSEA, and SRMSR with \texttt{CircE} \cite{Grassi2010a}. The dataset is available at \url{https://doi.org/10.21979/N9/9AZ21T} and the replication code is available at \url{https://github.com/ntudsp/satp-zsm-stage2}.

%% file: table/equipment.tex
\newcolumntype{Z}{>{\raggedright\arraybackslash}p{3.5cm}}
\newcolumntype{H}{>{\raggedright\arraybackslash}p{1.8cm}}
\newcolumntype{I}{>{\raggedright\arraybackslash}p{2.7cm}}
\newcolumntype{J}{>{\raggedright\arraybackslash}p{4.25cm}}
\newcolumntype{Y}{>{\raggedright\arraybackslash}p{4.25cm}}
\begin{table*}[tb]
\caption{Hardware specifications for the calibration and playback of binaural audio tracks in NTU and UPM}
\label{tab:equip}
\small
\begin{tabularx}{\linewidth}{lXX}
\toprule
Type                              
    & NTU   
    & UPM  

\\ \midrule

Headphones 
    & DT 990 Pro \newline 
    (Beyerdynamic GmbH \& Co.\ KG,\newline Heilbronn, Germany)
    & HD650 \newline
    (Sennheiser Electronic GmbH \& Co.\ KG,\newline Hanover, Germany)                    
\\ \midrule
Soundcard 
    & Ultralite AVB \newline
    (MOTU Inc, Cambridge, MA, USA)
    & UR22 \newline
    (Steinberg Media Technologies GmbH,\newline Hamburg, Germany)
\\ \midrule
Voltage meter
    & Fluke 79 Series II Multimeter \newline
    (Fluke Corporation, WA, USA)
    & CD800a \newline
    (Sanwa Denshi Co., Ltd, Osaka, Japan)
\\ \midrule
Acoustic environment
    & Listening room \newline
    (Noise Floor: \SI{30.2}{\decibelA} \cite{Ooi2023b})     
    & Recording studio \newline
    (Noise Floor: \SI{40.4}{\decibelA})                
\\
\bottomrule
\end{tabularx}
\end{table*}

%% file: table/demo.tex
\begin{table}

\caption{\label{tab:demo}Summary of reported demographic information by ethnonational groupings. Where applicable, standard deviations are shown in parentheses.}
\setlength{\tabcolsep}{3pt}
\centering
\begin{tabularx}{\linewidth}{lRRR}
\toprule
  & MY:M 
  & MY:O 
  & SG\\
  
\midrule
\textbf{Age} (years)
    & 24.00 (4.87) 
    & 23.09 (2.64) 
    & 25.91 (7.18)\\

\midrule
\textbf{Written Fluency} 
    & 8.74 (1.26) 
    & 6.38 (1.56) 
    & 8.06 (1.58)\\
\midrule
{\textbf{Spoken Fluency}}
    & 9.48 (0.72) 
    & 6.16 (1.53) 
    & Binary\\

\hspace{1em}Yes ($\ge\text{7}$) 
    & 31 
    & 17 
    & 32\\
\hspace{1em}No 
    & 0 
    & 15 
    & 0\\

\midrule
\multicolumn{4}{l}{\textbf{Gender}}\\
\hspace{1em}Female 
    & 15 
    & 16 
    & 16\\
    
\hspace{1em}Male 
    & 16 
    & 16 
    & 16\\
    
\bottomrule
\end{tabularx}
\end{table}

%% file: sections/03listening.tex
\section{Results: Listening experiment}\label{sec:listening}

For conciseness and readability, the stimuli are referenced in set notation when more than 5 individual stimuli are referenced in the text, i.e., $\mcsn$, $n\in\{1,2,\dots,26,27\}$, where $\mcsn$ is the $n$-th stimulus. Both the KWT and posthoc CIT for main-axis and derived-axis PAQ attributes, are detailed in \Cref{tab:kwt} and \Cref{tab:cit} in
\labelcref{sec:append_stats}, respectively.

The potential impact of order effects stemming from technical errors in the electronic form was first investigated. A two-sample KS test was conducted on each stimulus-attribute pair, comparing a group of 10 participants who were presented the stimuli in a fixed order to a group of 22 participants with the random order presentation, with BH correction for multiple comparisons. No significant differences between the fixed and random order groups were found across all stimulus-attribute pairs ($p$ > 0.05). These findings suggest that both groups exhibit similar distributions, indicating the absence of order effects. It is also worth noting that the KS test was performed only on the \scsg\ dataset as the participants in both the \scmym\ and \scmyo\ datasets were presented stimuli in a fixed order.

\subsection{Perceived affective quality attributes}

The KWT revealed significant differences between \scsg, \scmym\ and \scmyo\ in at least one PAQ attribute in 20 out of all $N=27$ stimuli, Hence, no differences were observed only in \SI{25.93}{\percent} of the stimuli, i.e. $\mcsn$, where $n \in \{9, 13, 19, 21, 23, 26, 27\}$.  Across the main axis attributes, significant differences were found in: \ie\ in \SI{18.52}{\percent} of $\mcsn$, where $n \in \{4,7,15,20,22\}$; \ip\ in \SI{22.22}{\percent} of $\mcsn$, where $n \in \{3,6,7,12,15,24\}$; \iu\ in \SI{14.81}{\percent} of $\mcsn$, where $n \in \{2,6,7,8\}$; and \ia\ in \SI{18.52}{\percent} of $\mcsn$, where $n \in \{3,6,7,11,20\}$. Among the derived axis attributes, significant differences were found in: \iv\ in \SI{37.04}{\percent} of $\mcsn$, where $n \in \{4,6,7,10,12,14,15,17,20,25\}$; \ica\ in \SI{25.93}{\percent} of $\mcsn$, where $n \in \{1,3,6,7,12,15,16\}$, \textit{monotonous} in \SI{18.52}{\percent} of $\mcsn$, where $n \in \{3,6,7,19,22\}$, and \ich\ in \SI{25.93}{\percent} of $\mcsn$, where $n \in \{4,6,7,10,12,16,22\}$. 

For significant KWT ($p<0.05$) with at least a small effect size ($\eta^2\ge0.01$), posthoc CIT was conducted on each attribute--stimulus pair to examine pairwise ethnonational differences between groups. Among main axis PAQ attributes, significant differences were found in \ie\ between \scmym\ and \scmyo\ in $\mcsn$, where $n \in \{7,20,22\}$ ($p_{\mathcal{S}_{20}},p_{\mathcal{S}_{22}}<0.05$ and $p_{\mathcal{S}_7}<0.01$); between \scsg\ and \scmym\ in $\mathcal{S}_{4}$ and $\mathcal{S}_{7}$ ($p_{\mathcal{S}_7}<0.01$; $p_{\mathcal{S}_4}<0.05$); and between \scsg\ and \scmyo\ in $\mathcal{S}_{15}$ ($p_{\mathcal{S}_{15}}<0.01$).

For \ip\, significant differences were found between \scmym\ and \scmyo\ in $\mathcal{S}_{3}$ ($p_{\mathcal{S}_3}<0.05$); between \scsg\ and \scmym\ in $\mathcal{S}_{6}$, $\mathcal{S}_{7}$ and $\mathcal{S}_{12}$ ($p_{\mathcal{S}_6}<0.0001$, $p_{\mathcal{S}_7}<0.01$, $p_{\mathcal{S}_{12}}<0.05$); and between \scsg\ and \scmyo\ in $\mathcal{S}_{3}$, $\mathcal{S}_6$, $\mathcal{S}_7$, and $\mathcal{S}_{24}$ ($p_{\mathcal{S}_3}<0.0001$; $p_{\mathcal{S}_7}<0.01$; $p_{\mathcal{S}_6}$,$p_{\mathcal{S}_{24}}<0.05$). 

\begin{figure}[ht]
    \centering
    \includegraphics[width=1\linewidth]{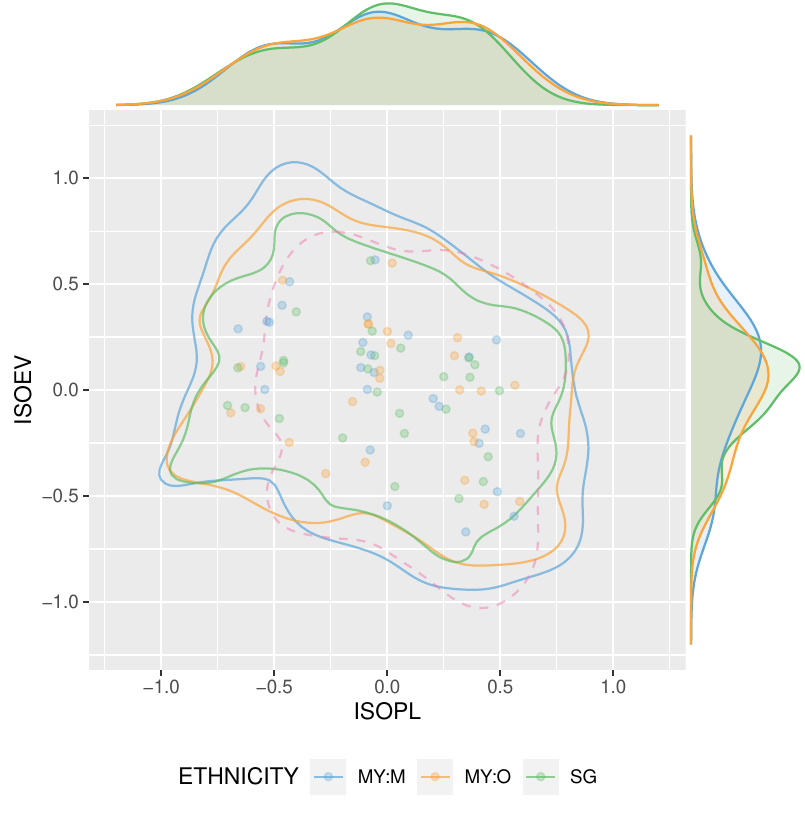}
    \caption{Median scores and \SI{20}{\percent} probability density contours of \iisopl\ and \iisoev\ of \scmym\ \mbox{(\textcolor{few_pal_1}{---})}, \scmyo\ \mbox{(\textcolor{few_pal_2}{---})}, and \scsg\ \mbox{(\textcolor{few_pal_3}{---})} across 27 stimuli. The KDE plots of the marginal \iisopl\ and \iisoev\ distributions are displayed along the respective axis margins. The \SI{20}{\percent} contour for ethnic Malays in the ARAUS dataset is indicated by the dashed magenta line \mbox{(\textcolor{few_pal_4}{- -})} \cite{Ooi2023a}.}
    \label{fig:isoplevcontall}
\end{figure}

In \iu\, significant differences were found between \scmym\ and \scmyo\ in $\mathcal{S}_{7}$ ($p_{\mathcal{S}_7}<0.05$); between \scsg\ and \scmym\ in $\mathcal{S}_{6}$ ($p_{\mathcal{S}_6}<0.01$); and between \scsg\ and \scmyo\ in $\mathcal{S}_{8}$ ($p_{\mathcal{S}_8}<0.05$). 

Lastly, for \ia, significant differences were found between \scmym\ and \scmyo\ in $\mathcal{S}_3$, ($p_{\mathcal{S}_3}<0.05$); between \scsg\ and \scmym\ in $\mathcal{S}_6$, $\mathcal{S}_7$, and $\mathcal{S}_{11}$ ($p_{\mathcal{S}_6}<0.001$, $p_{\mathcal{S}_7},p_{\mathcal{S}_{11}}<0.05$); and between \scsg\ and \scmyo\ in $\mathcal{S}_4$ ($p_{\mathcal{S}_4}<0.05$). 

Among derived PAQ attributes, significant differences were found in \iv\ between \scmym\ and \scmyo\ in $\mcsn$, where $n\in\{4,7,10,12,17,20,25\}$ ($p_{\mathcal{S}_{12}}<0.0001$; $p_{\mathcal{S}_7}, p_{\mathcal{S}_{10}}, p_{\mathcal{S}_{17}}<0.001$; $p_{\mathcal{S}_4},p_{\mathcal{S}_{20}},p_{\mathcal{S}_{25}}<0.05$); between \scsg\ and \scmym\ in $\mcsn$, where $n\in\{6,7,12,15,25\}$ ($p_{\mathcal{S}_{12}}<0.0001$; $p_{\mathcal{S}_7},p_{\mathcal{S}_{15}}<0.01$; $p_{\mathcal{S}_{6}},  p_{\mathcal{S}_{25}}<0.05$); and between \scsg\ and \scmyo\ only in $\mathcal{S}_{17}$ ($p_{\mathcal{S}_{17}}<0.05$). 

For \ica, significant differences were found between \scmym\ and \scmyo\ only in $\mathcal{S}_{15}$ ($p_{\mathcal{S}_{15}}<0.05$); between \scsg\ and \scmym\ in $\mathcal{S}_6$, $\mathcal{S}_7$, $\mathcal{S}_{12}$ and $\mathcal{S}_{16}$ ($p_{\mathcal{S}_6}<0.0001$, $p_{\mathcal{S}_{16}}<0.001$, $p_{\mathcal{S}_7}<0.01$, $p_{\mathcal{S}_{12}}<0.05$); and between \scsg\ and \scmyo\ in $\mcsn$, where $n\in\{1,3,7,6,15,24\}$ ($p_{\mathcal{S}_3},p_{\mathcal{S}_7}<0.01$; $p_{\mathcal{S}_1},p_{\mathcal{S}_{6}},p_{\mathcal{S}_{15}},p_{\mathcal{S}_{24}}<0.05$). 

In \im, significant differences were only found between \scsg\ and \scmym\ in $\mathcal{S}_3$ and $\mathcal{S}_7$ ($p_{\mathcal{S}_3}<0.01$; $p_{\mathcal{S}_7}<0.05$); and between \scsg\ and \scmyo\ in $\mathcal{S}_{6}$, $\mathcal{S}_7$ and $\mathcal{S}_{19}$ ($p_{\mathcal{S}_6}<0.001$; $p_{\mathcal{S}_7}<0.01$; $p_{\mathcal{S}_{19}}<0.05$). 

\begin{figure*}[ht]
    \centering
    \includegraphics[width=1\textwidth]{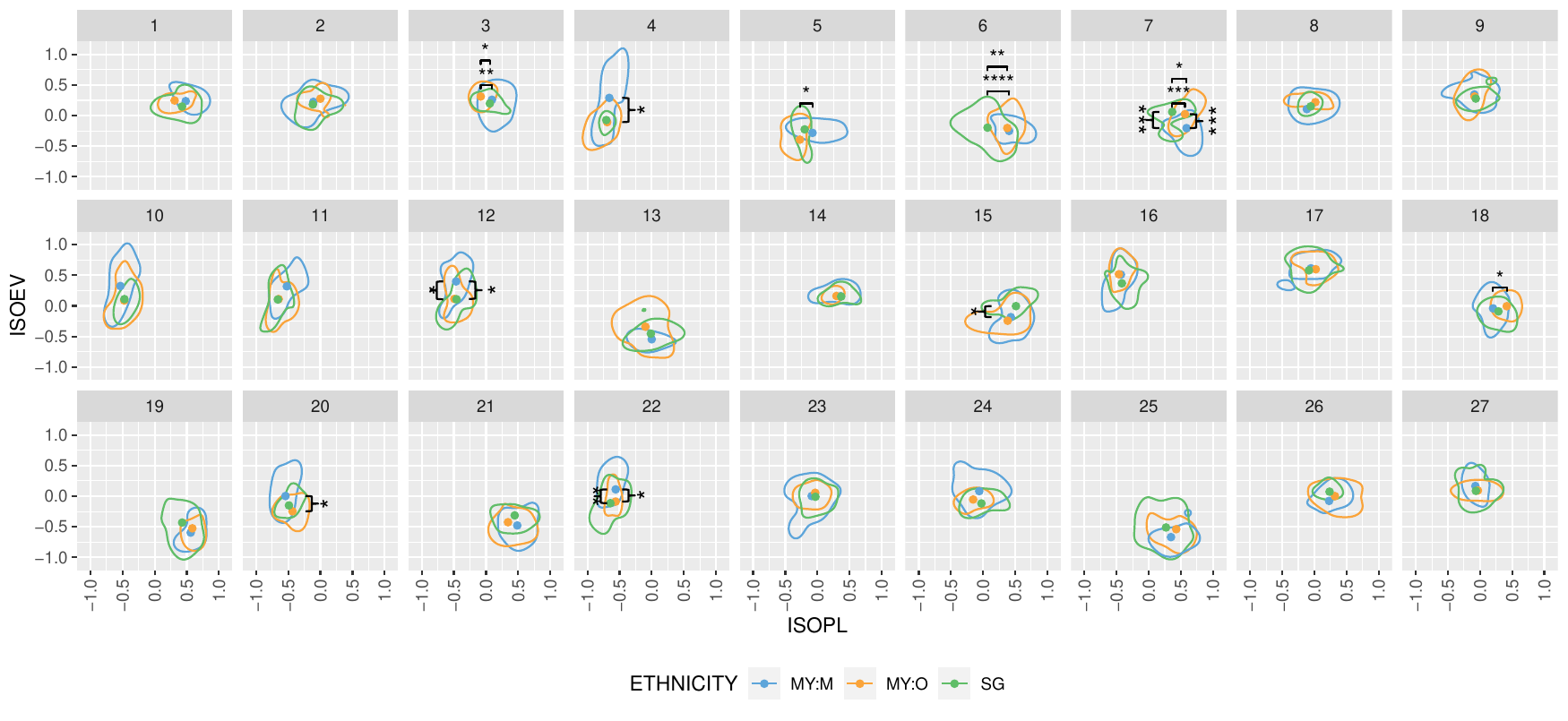}
    \caption{Median scores and contours of \iisopl\ and \iisoev\ of \scmym\ (\textcolor{few_pal_1}{---}), \scmyo\ (\textcolor{few_pal_2}{---}), and \scsg\ (\textcolor{few_pal_3}{---}) across 27 stimuli, wherein \SI{50}{\percent} of the individual scores lie within the median contours. Significant ethnonational differences in \iisopl\ are indicated by asterisks and horizontal squared braces, whereas those differences in \iisoev\ are marked with vertical curly braces.}
    \label{fig:isoplevcont}
\end{figure*}

Lastly, for \ich, significant differences were found between \scmym\ and \scmyo\ in $\mathcal{S}_4$, $\mathcal{S}_{10}$, $\mathcal{S}_{12}$ and $\mathcal{S}_{22}$ ($p_{\mathcal{S}_{12}}<0.01$; $p_{\mathcal{S}_4},p_{\mathcal{S}_{10}}$, $p_{\mathcal{S}_{22}}<0.05$); and between \scsg\ and \scmym\ in $\mathcal{S}_6$, $\mathcal{S}_7$ and $\mathcal{S}_{16}$ ($p_{\mathcal{S}_7}<0.01$; $p_{\mathcal{S}_{6}},p_{\mathcal{S}_{16}}<0.05$). The summary statistics and posthoc CIT significance levels across groups and stimuli for each PAQ attribute is plotted in \Cref{fig:boxplot}.

\subsection{ISO Pleasantness and ISO Eventfulness}

ISO 12913-3 describes a two-dimensional representation of soundscapes in \textit{Pleasantness} (\iisopl) and \textit{Eventfulness} (\iisoev) \cite{iso12913-3}, where \iisopl\ is obtained from the 8 PAQ attributes by
\begin{equation} \label{eqn:ISOPL}
    \frac{\iisopl}{k}=(p-a)+\cos\SI{45}{\degree}\cdot(\textit{ca}-\textit{ch}) + \cos\SI{45}{\degree}\cdot(v-m), 
\end{equation}
as a coordinate on the \textit{x}-axis, and \iisoev\ by
\begin{equation} \label{eqn:ISOEV}
    \frac{\iisoev}{k}=(e-u)+\cos\SI{45}{\degree}\cdot(\textit{ch}-\textit{ca}) + \cos\SI{45}{\degree}\cdot(v-m), 
\end{equation}
as a coordinate on the \textit{y}-axis. Here, $p, a, ca, ch, v, m, e$, and $u$ represent the ratings given for the PAQ attributes \ip, \ia, \ica, \ich, \iv, \im, \ie, and \iu, respectively. In addition, $k$ is a normalisation constant applied to constrain the values of \iisopl\ and \iisoev\ in the range $[-1,1]$. In ISO 12913-3, $k=\frac{1}{4}\left(1+\sqrt{2}\right)$ since the PAQ attributes lie in $\{1,2,3,4,5\}$. However, for this study, we use $k=\frac{1}{100}\left(1+\sqrt{2}\right)$ due to the 101-point scale for the PAQ attributes described in \Cref{sec:studyexptdesign}.

Median \iisopl\ and \iisoev\ scores across all participants for each stimuli across each group revealed a similar spread between ethnonational groups, as shown in \Cref{fig:isoplevcontall}. To further examine the distribution in the 2D soundscape space, \SI{20}{\percent} probability contours of the 2D kernel density estimates of all individual responses within each group were computed, which encircles most of the responses, i.e. at least \SI{20}{\percent} of the estimated density \cite{Mitchell2022HowData}. The participant responses exhibited a similar trend but differing degrees of spread in the \ica\ and \ica\ dimensions, where \scmym\ had the greatest spread, followed by \scmyo, then \scsg. Though exploratory, the contours indicate a compression in the \iv\ and \im\ dimensions in an otherwise acoustically diverse stimuli dataset \cite{Oberman2022}. 

Ethnonational differences in the \iisopl\ and \iisoev\ scores were investigated with the KWT, whereby significant differences were observed for \iisopl\ in $\mcsn$, $n\in\{3,5,7,18,19\}$ ($p_{\mathcal{S}_3}<0.001$; $p_{\mathcal{S}_7}<0.01$; $p_{\mathcal{S}_5},p_{\mathcal{S}_{18}},p_{\mathcal{S}_{19}}<0.05$), and for \iisoev\ in $\mcsn$, $n\in\{4,7,12,15,20,22\}$ ($p_{\mathcal{S}_7}<0.01$; $p_{\mathcal{S}_n}<0.05$, $n\in\{4,12,15,20,22\}$), as shown in \Cref{tab:kwt}.

Significant differences were observed in the posthoc CIT for \iisopl\ between \scmym\ and \scmyo\ in $\mathcal{S}_3$, $\mathcal{S}_5$ and $\mathcal{S}_{18}$ ($p_{\mathcal{S}_3}<0.001$; $p_{\mathcal{S}_5}<0.01$; $p_{\mathcal{S}_{18}}<0.05$); between \scsg\ and \scmym\ only in $\mathcal{S}_7$ ($p_{\mathcal{S}_7}<0.01$); and between \scsg\ and \scmyo\ in $\mathcal{S}_3$, $\mathcal{S}_7$, and $\mathcal{S}_{19}$ ($p_{\mathcal{S}_3}<0.01$; $p_{\mathcal{S}_7}<0.001$; $p_{\mathcal{S}_{19}}<0.05$). For \iisoev, significant differences were found in posthoc CIT only between \scmym\ and \scmyo\ in $\mathcal{S}_4$, $\mathcal{S}_7$, $\mathcal{S}_{12}$ and $\mathcal{S}_{22}$ ($p_{\mathcal{S}_4},p_{\mathcal{S}_{12}},p_{\mathcal{S}_{22}}<0.05$; $p_{\mathcal{S}_7}<0.01$); and between \scsg\ and \scmym\ in $\mathcal{S}_7$ and $\mathcal{S}_{15}$ ($p_{\mathcal{S}_7}<0.05$; $p_{\mathcal{S}_{15}}<0.01$). The detailed results of the posthoc CIT for significant cases in the KWT is summarised in \Cref{tab:cit}.

To visualise the ethnonational differences in \iisopl\ and \iisoev\ in the two-dimensional space, the distribution of \iisopl\ and \iisoev\ scores from individual participants were computed as median contours across ethnonational groups for each stimulus, as shown in \Cref{fig:isoplevcont}. It is thus evident from \Cref{fig:isoplevcont} that only in $\mathcal{S}_7$ that both \iisopl\ and \iisoev\ differed significantly across all three groups. 

\subsection{Principal components analysis} \label{sec:PCA}

Kaiser-Meyer-Olkin (KMO) and Bartlett tests of sphericity were conducted to first determine the suitability of the data for Principal Components Analysis (PCA). The KMO tests revealed borderline adequacy of the sample size and variable intercorrelations, where the overall measure of sampling adequacy was 0.815, 0.754, and 0.770 for \scmym, \scmyo, and \scsg, respectively. The Bartlett tests of sphericity showed significant differences across all groups ($p<0.0001$), further signaling the suitability for PCA. 

\begin{figure*}[ht]
    \centering
    \includegraphics[width=1\textwidth]{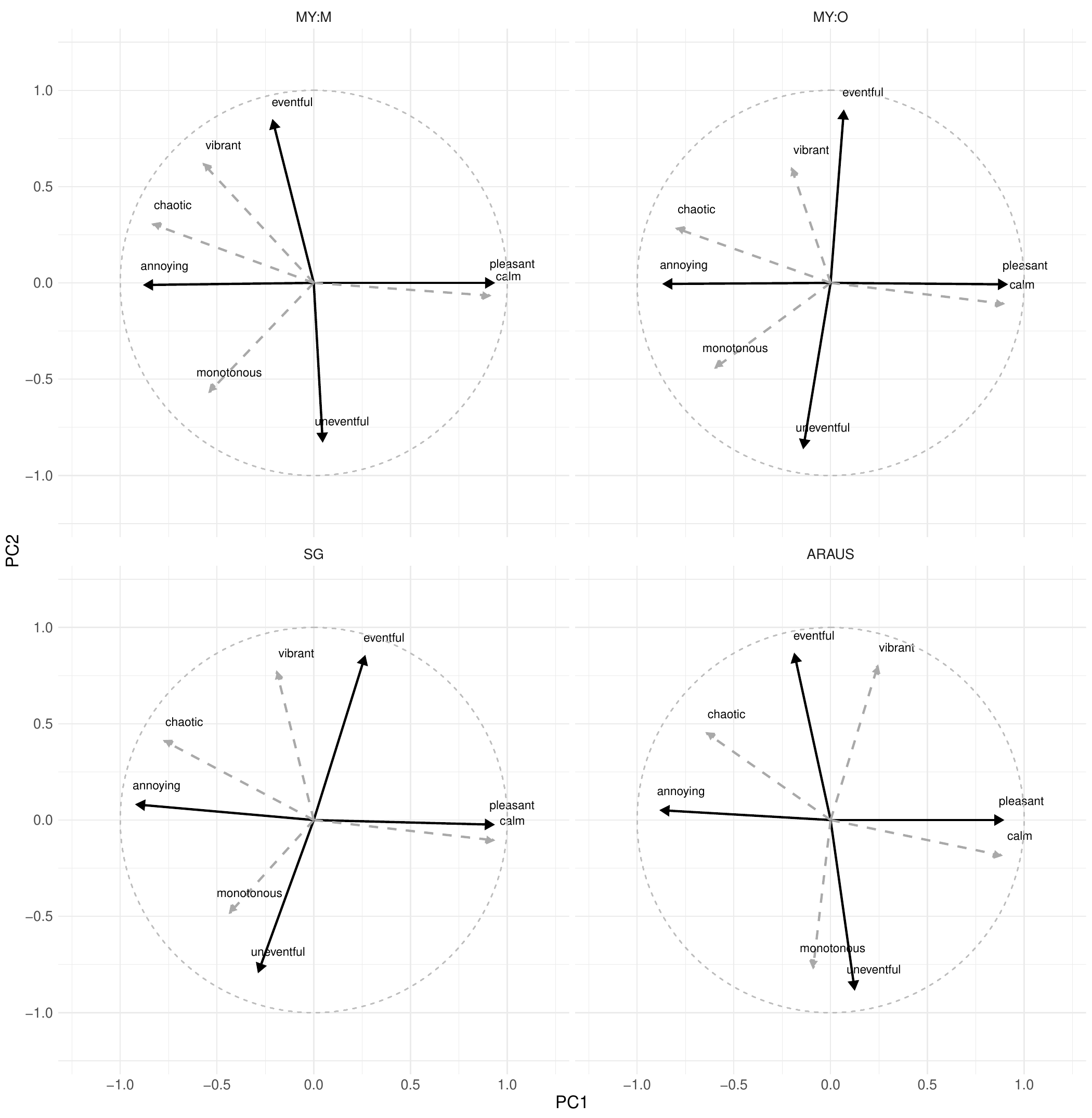}
    \caption{The 8 perceived affective quality attributes on the 2D principal component space for \scsg, \scmym\ and \scmyo\ groups. Solid and dashed lines represent the main and derived axis attributes from ISO 12913-3, respectively.}
    \label{fig:pcaproj}
\end{figure*}

The PCA revealed that the first two principal components (PC) correspond to the \ip--\ia\ (PC1) and \ie--\iu\ (PC2) axes, which accounted for \SI{74.31}{\percent} (PC1: \SI{49.44}{\percent}; PC2: \SI{24.87}{\percent}), \SI{69.25}{\percent} (PC1: \SI{42.39}{\percent}; PC2: \SI{26.85}{\percent}), and \SI{72.77}{\percent} (PC1: \SI{43.79}{\percent}; PC2: \SI{28.98}{\percent}) of the total explained variance for \scmym, \scmyo, and \scsg, respectively.

Component loadings of the 8 PAQ attributes across \scmym, \scmyo, and \scsg\ groups are plotted onto the two-dimensional (2D) PC1--PC2 space, where the attributes corresponding to the main axes in ISO 12913-3 are indicated in solid lines and the derived axes are indicated in dashed lines, as shown in \Cref{fig:pcaproj}. The component loadings mostly resemble the order in ISO 12913-3 circumplex PAQ model after adjustments (reflection about the y-axis for \scmym\ and \scsg). Owing to the stability of the \ia-\ip\ dimension, PCA loadings of the \scmym group were rotated to align the \ia-\ip\ dimension (PC1) to the \textit{x}-axis.

Both the \textit{Pleasantness} (PC1) and \textit{Eventfulness} (PC2) axes appear roughly orthogonal in all groups, but skewed counter-clockwise by about \SI{20}{\degree} in \scmym. After accounting for the skew, the distribution of derived axis attributes appear similar. A large deviation was observed across all groups in \iv\ and \ica, where \iv\ leaned towards \ia\ rather than \ip; and \ica\ appears to be collapsing to \ip, which signals a high correlation between the variables.

%% file: sections/04circumplex.tex
\section{Circumplexity analysis} \label{sec:circum}

Although not explicitly stated in ISO 12913-3, the two-dimensional PAQ model (i.e. Figure A.1 in \cite{iso12913-3}) takes reference from Axelsson's proposed ``circumplex'' model for PAQ \cite{Axelsson2010}, and Russell's circumplex model of affect \cite{Russell1980AAffect.}, as mentioned in \Cref{sec:intro-circ}. Prior to assessing circumplexity, it is important to identify which specific circular model and its underlying assumptions that are applicable to the PAQ model. 

\subsection{Circulant and quasi-circumplex models} \label{sec:circquasi}
\citet{Guttman1954} presented two specific models of circumplexity: the \textit{circulant} model, characterized by variables are equally spaced around the circle (equal angles) and equidistant from the centre of the circle (equal communality); and the \textit{quasi-circumplex} model, where variables are arranged in a circle but without equal angles or communality \cite{Guttman1954,Tracey2000}. The quasi-circumplex model could be further conditioned on each of the circulant conditions to yield four circumplex models: (1) circulant, (2) quasi-circumplex with equal communality (EC), (3) quasi-circumplex with equal angles (EA), and (4) quasi-circumplex. The adherence to at least the quasi-circumplex with equal angles is crucial for validating the assumption underlying the attributes of \iisopl\ and \iisoev, which are computed based on a \SI{45}{\degree} separation between adjacent variables as specified in ISO/TS~12913-3. 

To assess the four circumplex models, a confirmatory covariance structure modelling technique \cite{Assumptions1992,Gurtman2003,Tracey2000}, has been developed and is available as a DOS computer program \texttt{CIRCUM} \cite{Assumptions1992}, and subsequently implemented as an R-package \texttt{CircE} \cite{Grassi2010a}. The degree of fit to each model is evaluated using three indices and their recommended thresholds: Comparative Fit Index ($\textit{CFI}\ge0.90$) \cite{Gurtman2003}, Root Mean Square Error Approximation ($\textit{RMSEA}\le0.13$) \cite{Gurtman2003}, and the Standardised Root Mean Square Residual ($\textit{SRMSR}<0.08$) \cite{Hu1999,Moshona2023}. The \texttt{CircE} program examines the correlation matrix between the variables (e.g. Table 22.1 in \cite{Tracey2000}, Table 5 in \cite{Lam2022c}). For the circulant model, the inequality requirement of $P_1>P_2>P_3>P_4$ should be met (i.e. Table 22.2 in \cite{Tracey2000}). This implies that correlations between adjacent variables ($P_1$) should be greater than orthogonal variable correlations ($P_2$), $P_2$ should be greater than correlations of variables \SI{135}{\degree} apart ($P_3$), and $P_3$ should be greater than correlations of opposing variables on each axis ($P_4$). Additionally, the circulant model explicitly requires the relations at similar distances to be equivalent, for example, all $P_1$ should be equivalent.

The 8-attribute PAQ model consists of a total of 28 correlation pairs, computed using Pearson's method \cite{Lam2022c}. The fit index \textit{CFI} was found to be above \num{0.90} across all groups in the quasi-circumplex model; in only the \scmym\ and \scmyo\ in the EC quasi-circumplex model; indicating a good fit. However, in the EA quasi-circumplex and circulant models, the \textit{CFI} values were below \num{0.90}, indcating a poor fit. The RMSEA estimates further revealed that the only the quasi-circumplex across all groups and the EC quasi-circumplex model in \scmym\ provided good approximations of the data, with values $\le0.13$. Regarding the SRMSR estimates, only the quasi-circumplex model was a good fit across all groups, with values $<0.06$. Therefore, based on the remaining the fit indices, it can be concluded that the data neither fits into the circulant nor the EA circumplex models. 

\input{table/modelfitnew_araus}

\subsection{Circular order model}

Another confirmatory approach to assess the adequacy of the correlation matrix's fit to a circumplex model is through the utilization of a randomization test of hypothesized order relations (RTHOR; \cite{Hubert1987,Tracey2000,Locke2019}). The RTHOR assesses the fit to the circular order model, which shares similarities with the circulant model but does not explicitly test for the equality of relations and spacing \cite{Tracey2000}. Like the circulant model, the circular order model requires strict adherence to the inequality order of correlations between variables. The correspondence index (CI) is used to evaluate the circular model fit of the correlation matrices across groups, with a scale ranging from -1 (complete violation) to 0 (chance) to 1.0 (perfect fit), as proposed by \citet{Tracey2000}. A \textit{CI} score of 0.5 indicates that \SI{75}{\percent} of the predictions met the inequality criteria. In total, 288 predictions were tested for violation \cite{Zeigler-Hill2010,Locke2019}, where \SI{79.86}{\percent} of the predictions conformed to the inequality criteria in \scmym\ ($\textit{CI}=0.597$, $p<0.01$), \SI{84.03}{\percent} in \scmyo\ ($\textit{CI}=0.681$, $p<0.01$), and \SI{85.30}{\percent} in \scsg\ ($\textit{CI}=0.706$, $p<0.01$), as shown in \Cref{tab:modelfitnew_araus}. 

\subsection{Structural summary method}

In addition to the model fit indices \textit{CFI}, \textit{RMSEA}, and \textit{SRMSR} for equal spacing and communality, the overall circulant conformity of the underlying circumplexity could be further examined via the structural summary method (SSM) \cite{Zimmermann2017}, where the sinusoidal fit of the correlations between variables are examined \cite{Zimmermann2017,Locke2019}.  The poor SSM sinusoidal model fit ($<0.7$) across \scmym\ ($R^2=0.395$) and \scmyo\ ($R^2=0.637$), and marginally adequate ($R^2\ge0.7$) but short of a good fit ($R^2\ge0.8$) in \scsg\ ($R^2=0.706$). further affirms that the translated \sczsm\ PAQ model does not meet the circulant circumplexity requirements [i.e. equal communality (radii) and spacing].

%% file: table/modelfitnew_araus.tex
\begin{table*}

\caption{\label{tab:modelfitnew_araus}Summary of model fitting indexes, where conformity to the recommended threshold is indicated in \textbf{bold}}
\centering
\begin{tabularx}{\textwidth}{{>{\raggedright\arraybackslash}p{2cm}}X*{4}{>{\centering\arraybackslash}p{2cm}}}
\toprule
  &  
  & MY:M 
  & MY:O 
  & SG
  & ARAUS \\
  
\midrule
\multicolumn{5}{l}{\textit{Circular order}} \\
CI ($p$) 
&  
& 0.597 (0.002) 
& 0.681 (0.002) 
& 0.701 (0.002) 
& 0.847 (0.000) \\

\midrule
\multicolumn{5}{l}{\textit{Quasi-circumplex}} \\
CFI 
& $\ge0.90$ 
& \textbf{1.000} 
& \textbf{1.000} 
& \textbf{1.000}
& \textbf{1.000} \\

RMSEA 
& $\le0.13$ 
& \textbf{0.000} 
& \textbf{0.000} 
& \textbf{0.000}
& \textbf{0.000} \\

SRMSR 
& $<0.06$ 
& \textbf{0.016} 
& \textbf{0.020} 
& \textbf{0.037}
& \textbf{0.023} \\

\midrule
\multicolumn{5}{l}{\textit{Equal angles}} \\
CFI 
& $\ge0.90$ 
& 0.734 
& 0.825 
& 0.656
& \textbf{0.949}\\

RMSEA 
& $\le0.13$ 
& 0.254 
& 0.185 
& 0.293
& \textbf{0.097}\\

SRMSR 
& $<0.06$ 
& 0.254 
& 0.217 
& 0.231
& 0.159\\

\midrule
\multicolumn{5}{l}{\textit{Equal communality}} \\
CFI 
& $\ge0.90$ 
& \textbf{0.992} 
& \textbf{0.907} 
& 0.881 
& \textbf{1.000} \\

RMSEA 
& $\le0.13$ 
& \textbf{0.044} 
& 0.134 
& 0.172
& \textbf{0.000} \\

SRMSR 
& $<0.06$ 
& 0.114 
& 0.108 
& 0.142
& \textbf{0.056}\\

\midrule
\multicolumn{5}{l}{\textit{Equal communality and angles}} \\
CFI 
& $\ge0.90$ 
& 0.543 
& 0.611 
& 0.486
& \textbf{0.965} \\

RMSEA
& $\le0.13$ 
& 0.282 
& 0.234 
& 0.305
& \textbf{0.068} \\

SRMSR 
& $<0.06$ 
& 0.307 
& 0.237 
& 0.269
& 0.168 \\

\midrule
\multicolumn{5}{l}{\textit{Circulant (Equal communality and angles)}} \\
SSM 
& $\ge 0.7$ 
& 0.395 
& 0.637 
& \textbf{0.706}
& \textbf{0.803} \\

\bottomrule
\end{tabularx}
\end{table*}

%% file: sections/05discussion.tex
\section{Discussion}\label{sec:discussion}

The following discussion addresses the research questions posed in \Cref{sec:rq}. Firstly, the effect of ethnonational differences on the PAQ attributes is examined in \Cref{sec:dis-ethno}. Subsequently, the adherence of the PAQ attributes to the circumplex model is discussed in \Cref{sec:dis-circum}. Next, the alignment between Stage 2 listening test and quantitative method attributes from Stage 1 is investigated in \Cref{sec:dis-stage1}. Lastly, the limitations of this study and future research directions are discussed in \Cref{sec:dis-limit}.

\subsection{Effect of ethnonational differences on perceived affective quality attributes in Bahasa Melayu} \label{sec:dis-ethno}

The \scmyo\ PAQ responses exhibited the highest overall similarity to \scsg, with only 16 out of 270 attributes (\SI{5.93}{\percent}) showing significant differences. In comparison, \scmym\ and \scmyo\ had differences in 22 out of 270 attributes (\SI{8.15}{\percent}), while  the greatest dissimilarities occurred between \scmym\ and \scsg, with 31 out of the 270 attributes (\SI{11.48}{\percent}) being significantly different. Hence, the soundscape perception of Malaysian \sczsm\ speakers from other non-native ethnicities appears to be closer to that of native ethnic Malays in Singapore than native ethnic Malays Malaysia.

These disparities between \scsg\ and either \scmy\ groups (\scmym, \scmyo) can be attributed to the evaluation of stimuli $\cals_6$ and $\cals_7$. Among the significant differences between \scsg\ and \scmym, \num{17} of the \num{31} attributes (\SI{54.44}{\percent}) were related to $\cals_6$ and $\cals_7$. Similarly, between \scsg\ and \scmyo, \num{9} out of the \num{16} attributes (\SI{56.25}{\percent}) that showed significant differences were from $\cals_6$ and $\cals_7$. When compared \scmym\ or \scmyo, \scsg\ had significantly lower scores for \ip, \ica, and \iisopl\ in both $\cals_6$ and $\cals_7$. Consequently, the \im\ scores were significantly higher in \scsg\ compared to \scmym\ or \scmyo. However, significant differences in \iv, \iu, \ia\ and \ich\ scores were only observed between \scsg\ and \scmym. 

While both $\cals_6$ (\texttt{E02}) and $\cals_7$ (\texttt{E05}) contained nature sounds, $\cals_6$ predominantly featured fairly loud and continuous water sounds (\SI{67.4}{\decibelA}), whereas $\cals_7$ depicted a quiet seaside scene dominated by human voices (\SI{55.06}{\decibelA}) \cite{Lam2022ICSV,MonteiroAntunes2023}. These variations between \scsg\ and \scmy\ groups may stem from cultural and geographical variations. Singapore, as a densely populated and highly developed urban environment, has limited access to water features with vigorous flow and tranquil waterfronts, which are more prevalent in Malaysia. Notably, no differences in PAQ were found in the only other stimulus with water sounds ($\cals_{21}$), providing further evidence that the dominant soft variable water sounds in $\cals_{21}$ are generally perceived as pleasant \cite{RadstenEkman2015a,Hong2020b}.

Additionally, the differences between \scmym\ and either \scmyo\ and \scsg\ can be attributed to variations in \iv\ scores. This is evident from the PCA loadings in \Cref{fig:pcaproj}, where the \iv\ axis was further skewed towards the \ia\ direction in \scmym\ responses. Excluding variables in $\cals_6$ and $\cals_7$, as well as the \iisopl\ and \iisoev\ scores, \iv\ scores accounted for 4 out of 11 attribute differences (\SI{36.36}{\percent}) in the comparison between \scmym\ and \scsg. Similarly, in the comparison between \scmym\ and \scmyo, 5 out of 14 attribute differences were due to differences in \iv\ scores (\SI{35.71}{\percent}). Notably, there were no significant differences in \iv\ scores across all stimuli between \scmyo\ and \scsg.

Given that significant differences were observed only in \ia\ in $\cals_{11}$ (\texttt{E12b}) between \scmym\ and \scsg, and in $\cals_3$ (\texttt{CG07}) between \scmym\ and \scmyo, it can be inferred that the adoption of region-specific translations such as \imembi\ for Singapore and \imenje\ for Malaysia, appropriately addressed cultural distinctions.

\subsection{Circumplexity of the perceived affective quality attributes in Bahasa Melayu} \label{sec:dis-circum}

The circumplexity tests conducted in \Cref{sec:circum} align with SATP datasets from Germany (ISO-639:3:\textsc{deu}) and the UK (ISO 639:3:\textsc{eng}) revealing that the \sczsm\ PAQ data only conforms to the quasi-circumplex model, while to meet the desired equal angle model \cite{Moshona2023}. These findings raise concerns regarding the validity of the \iisopl\ and \iisoev\ attributes when the fundamental assumption of equal angle representation in the circumplex model is violated. 

Nevertheless, the deviation observed in the circumplex model resulting from \iv\ and \ie\ scores, which was also observed in the \scspa\ study \cite{Vida2023}, can plausibly be attributed to the respondents' unfamiliarity with the UK-centric context of the SATP stimuli dataset. This tendency is evident in the responses of \num{29} of \num{600} participants who identified themselves as ethnic Malay in the ARAUS dataset \cite{Ooi2023a}. In particular, the PAQ scores demonstrated a strong fit with almost all the fit indices of the model, as presented in \Cref{tab:modelfitnew_araus}. Each of the \num{29} participants evaluated seven \num{30}-\si{\second} excerpts, randomly selected from a pool of \num{234} unique \num{1}-\si{\minute} long soundscape recordings sourced from the Urban Soundcapes of the World (USotW) dataset \cite{DeCoensel2017UrbanMind}. Each \num{1}-\si{\minute} recording was divided into two equal \num{30}-\si{\second} excerpts. Consequently, the total evaluation comprised \num{149} \num{30}-\si{\second} excerpts from \num{65} distinct \num{1}-\si{\minute} soundscape recordings. It is important to note that the USotW dataset encompasses recordings from various major cities around the world, selected based on a site selection protocol to ensure a diverse representation within the PAQ circumplex \cite{DeCoensel2017UrbanMind}. Nonetheless, the questionnaire in ARAUS was conducted in \sceng\ and does not fully account for possible linguistic differences affecting the circumplexity in the \sczsm\ responses.

\begin{figure}
    \centering
    \includegraphics[width=1\linewidth]{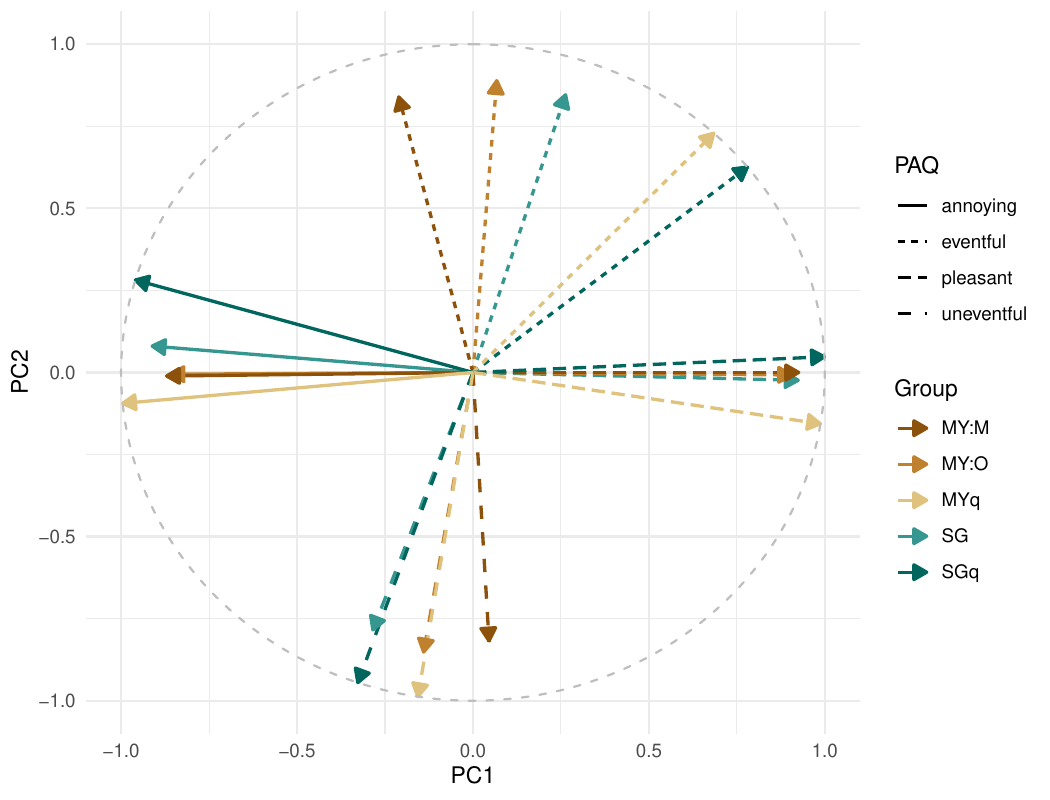}
    \caption{Main-axis PAQ PCA loadings on the 2D principal component space for \scmym\ (\textcolor{brbg_pal_1}{---}), \scmyo\ (\textcolor{brbg_pal_2}{---}) and \scsg\ (\textcolor{brbg_pal_8}{---}) groups from Stage 2; and the estimated main-axis loadings for  \textsc{myq} (\textcolor{brbg_pal_3}{---}) and \textsc{sgq} (\textcolor{brbg_pal_9}{---}) from Stage 1.}
    \label{fig:pcaest}
\end{figure}

\subsection{Alignment between stage 2 listening tests and quantitative method attributes from stage 1} \label{sec:dis-stage1}

Quantitative evaluation of SATP Stage 1 translations to \sczsm also revealed potential circumplexity violations. Particularly, in the main-axis attributes, \imeriah\ (\ie) exhibited low orthogonality (\scortho) towards \ia\ and \ip, and low non-connotativeness (\scncon) and implicative balance (\scibal) towards \iv\ and \ich. To aid in the visual comparision, bias scores (i.e. $r_\text{bias}$) used to examine the \scortho\ of each main-axis attribute was translated to estimated angular displacements in the clockwise direction from the ideal axial positions on the PCA loadings plot, given by
\begin{equation}
\Delta{\theta_\textit{paq}}=\frac{r_\textit{bias,paq}}{10}\times\SI{180}{\degree}-\SI{90}{\degree},
\end{equation}

\noindent where $\textit{paq}\in\{\ie, \ip, \iu, \ia\}$. Hence, $\Delta{\theta_\ie}=\SI{20}{\degree}$ indicates a \SI{20}{\degree} deviation from the y-axis in the clockwise direction towards \ip. The estimated angular displacements of the attributes of the main axis in the Singaporean (\scsgq) and Malaysian (\scmyq) groups from the quantitative survey in Stage 1 \cite{Lam2022MalayP1}, are depicted in \Cref{fig:pcaest}. 

From the $\Delta{\theta_\ie}$, it is evident that the low \scibal\ scores of \imeriah\ (\ie) arises from the skew toward \ip\ in Stage 1, which is also reflected in the \imeriah\ (\ie) loadings across \scmyo\ and \scsg\ in Stage 2. However, \imeriah\ (\ie) and \itimer\ (\iu) from \scmym\ deviated greatly from $\Delta{\theta_\ie}$ and $\Delta{\theta_\iu}$, where \imeriah\ biased toward \ia\ instead of \ip, and \iu\ biased toward \ip\ instead of \ia. 

Main axis attribute biases across \scmyo\ and \scsg\ in Stage 2 were aligned with the respective \scmyq\ and \scsgq\ biases in Stage 1. The observed discrepancies in the \ie--\iu\ dimension could be attributed to the objective nature of the dimension in terms of the perceived activity levels \cite{Nagahata2022}, which could be influenced by landscape morphology and culture \cite{Papadakis2022}. For instance, the greater bias of \imeriah\ (\ie) towards \ip\ in \scsg\ may be greatly influenced by the significantly higher population density in Singapore than most of Malaysia. Appropriately, a lack of activity could be associated with greater annoyance in Singapore. 

Among derived-axis attributes (\iv, \ica, \im, \ich), the PCA loadings were visually similar across all ethnonational groups, as shown in \Cref{fig:pcaproj}. Although \itenang\ (\ica) was found to strongly imply \ip\ in Stage 1 across both \scmyq\ and \scsgq, \itenang\ (\ica) and \imenye\ (\ip) became almost synonymous across all ethnonational groups in the listening tests. Notably, \irancak\ (\iv) exhibited stronger associations (i.e. $r_\textit{assoc}$) and implications $r_\textit{impl}$ towards \ie\ than \ip\ in Stage 1, but completely diverged into the \ich\ quadrant in Stage 2 listening tests, as shown in \Cref{fig:pcaproj}. It is worth noting that \textsc{orth} (i.e. $r_\textit{bias}$) scores were omitted for derived-axis attributes in Stage 1, and thus comparisons cannot be made. 

Besides possible inherent variability in the non-emotive nature of the \ie--\iu\ dimension \cite{Nagahata2022}, the lack of context due to the absence of stimuli during the quantitative evaluation in Stage 1 is another potential confounding factor.

\subsection{Limitations and future work} \label{sec:dis-limit}

Despite the valuable insights gained from this study, several limitations should be acknowledged, providing opportunities for future research to further enhance our understanding of PAQ attributes in cross-cultural and cross-national contexts, as well as in direct translations to other languages.

The SATP dataset listening test protocol prioritized inclusiveness and replicability over ecological validity by omitting the accompanying visuals to the binaural audio stimuli. Although the audio stimuli plays a dominant role in the perception of urban soundscapes, the audiovisual interaction effect and potential lack of context in such cross-cultural evaluation should not be ignored \cite{Li2020}. Additionally, the low-cost audio reproduction calibration procedure in the SATP listening test protocol had been shown to significantly influence the perceptual responses to the stumili as compared to a standard calibration procedure for faithful reproduction using a calibrated head and torso simulator \cite{Lam2022ICSV,Lam2022ICA}. Hence, follow-up studies to investigate the effectiveness of the translated PAQ attributes should consider audiovisual reproduction protocols with high ecological validity.

Another limitation pertains to the stimuli used in the study. Although efforts were made to include a diverse range of soundscapes, the UK-centric selection may still not fully represent the entire spectrum of soundscape environments. To enhance the comprehensiveness of the analysis of soundscape perception, future research should incorporate a more diverse range of soundscapes to evoke a wider range of responses in the PAQ circumplex.

The full characterisation of the ethnonational differences may have been restricted by the limited demographic diversity in this study. The relatively young demographic in \scsg\ could only be representative to the \iseba\ speaking demographic of the entire \sczsm-speaking community in \scsg, as described in \Cref{sec:evolutionBM}. Hence, including the non \iseba\ speaking population in \scsg\ would provide a more comprehensive understanding of the PAQ attributes in the \sczsm-speaking community.

Deviations in certain PAQ attributes to other translated languages could also be attributed to linguistic differences and limitations of the bilingual approach in the translation methodology. The interpretation of the PAQ attributes in \sceng\ could be influenced by the indigenised English in Singapore and Malaysia, as well as the possible constrained effects of the translation to \sczsm\ by bilingual individuals \cite{Kruger2016}. Moreover, certain affective terms may not exist in certain languages or may not be emotive or affective in nature \cite{Ortony2022}. Hence, it is important to determine if perceptual differences exist between English-speaking cultures in the PAQ attributes before conducting further translations.

%% file: sections/06conclusion.tex
\section{Conclusion}\label{sec:conclusion}

This work aims to validate the translated perceived affective quality (PAQ) attributes in ISO 639-3:\sczsm\ through a common listening test protocol under the Soundscape Attributes Translation Project (SATP) initiative. The influence of \sczsm\ proficiency on soundscape evaluations is investigated among \sczsm\ speakers across Singapore (\scsg) and Malaysia (\scmy): native Singaporean ethnic Malays (\scsg), native Malaysian ethnic Malays (\scmym), and other Malaysian ethnicity (\scmyo).

Firstly, significant differences were observed across ethnonational groups and the 27 acoustically diverse soundscapes from the SATP initiative. The \sgmyo\ group exhibited the highest overall similarity to \scsg, with only \SI{5.93}{\percent} of the PAQ attributes across the 27 stimuli showing significant differences. On the other hand, \scmym\ and \scsg\ demonstrated the greatest dissimilarities, with \SI{11.48}{\percent} of the attributes showing significant differences, highlighting the influence of ethnonational factors on soundscape perception. These disparities can be attributed to variations in the \iv\ scores, which accounted for most of the disparities between \scmym\ and either \scmyo\ or \scsg\ groups. Additionally, no difference were observed between \scmym\ and \scsg\ groups in the \iv\ scores.

The evaluation of specific stimuli, particularly S6 and S7, was also observed to elicited significant differences between \scsg\ and both \scmy\ groups. Cultural and geographical variations was postulated to play a role in these differences, as Singapore's urban environment, with limited access to certain natural elements like vigorous water flow and tranquil waterfronts, contrasted with the more prevalent presence of such elements in Malaysia.

Furthermore, the circumplexity tests revealed that the \sczsm\ PAQ data only conformed to the quasi-circumplex model, raising concerns about the validity of the \iisopl\ and \iisoev\ attributes when the equal angle representation assumption is violated. However, the deviation observed in the circumplex model, particularly influenced by vibrant and eventful scores, can be plausibly attributed to respondents' unfamiliarity with the UK-centric context of the SATP stimuli dataset.

Finally, the alignment between Stage 2 listening tests and quantitative method from Stage 1 further supported the presence of circumplexity violations. Particularly, the skew in the \ie--\iu\ dimension but could not support the variations observed in the \iv\ attribute due to the absence of orthogonality scores for \iv\ in Stage 1. The discrepancies observed in the \ie--\iu\ dimension may be influenced by the objective nature of perceived activity levels, landscape morphology, and cultural factors.

These findings provide valuable insights into the ethnonational differences in the \sczsm\ PAQ attributes and contribute to the understanding of soundscape perception across \sczsm--speaking populations of different ethnicities and geographies. Future research should continue to explore contextual aspects to enhance the accuracy and applicability of the \sczsm\ PAQ attributes, and investigate the cultural and geographical aspects that underpin the perception of the \ie--\iu\ dimension.

%% file: sections/99endmatter.tex
\section*{Data Availability}
The data that support the findings of this study are openly available in NTU research data repository DR-NTU (Data) at \url{https://doi.org/10.21979/N9/9AZ21T}, and replication code used in this study is available on GitHub at the following repository: \url{https://github.com/ntudsp/satp-zsm-stage2}. The code includes all the necessary scripts, functions, and instructions to reproduce the results reported in the study. Except for restricted access to the SATP stimuli dataset hosted at \url{https://doi.org/10.5281/zenodo.6914434}, the ARAUS dataset and data from \citet{Lam2022MalayP1} referenced in the paper are openly available in NTU research data repository DR-NTU (Data) at \url{https://doi.org/10.21979/N9/9OTEVX} and \url{https://doi.org/10.21979/N9/0NE37R}, respectively.

\section*{Declaration of competing interest}
The authors declare that they have no known competing financial interests or personal relationships that could have appeared to influence the work reported in this paper.

\section*{Acknowledgments}
This work was supported by the National Research Foundation, Singapore, and Ministry of National Development, Singapore under the Cities of Tomorrow R\&D Program (CoT Award: COT-V4-2020-1). Any opinions, findings and conclusions or recommendations expressed in this material are those of the authors and do not reflect the view of National Research Foundation, Singapore, and Ministry of National Development, Singapore.

The authors would like to thank Dr.\ Francesco Aletta, Dr.\ Tin Oberman, Dr.\ Andrew Mitchell, and Prof.\ Jian Kang, of the UCL Institute for Environmental Design and Engineering, The Bartlett Faculty of the Built Environment, University College London (UCL), London, United Kingdom, for coordinating the SATP project and providing assistance for the Bahasa Melayu Working Group.

\printcredits


\bibliographystyle{model1-num-names}

\bibliography{refbhan}

%% file: sections/98appendix.tex


\input{commands/appendix-start}
\onecolumn
\section{Translated perceived affective quality questionnaire in ISO~639-3:\textsc{zsm}}

\input{table/inittransPAQ}

\clearpage
\input{commands/next-appendix}


\section{Results of statistical tests on the evaluation scores} \label{sec:append_stats}

\setcounter{table}{0}
\input{table/kwtTable}
\input{table/citTable}

\clearpage
\input{commands/next-appendix}
\section{Summary statistics}

\setcounter{figure}{0}
\begin{figure*}[h]
    \centering
    \includegraphics[width=1\textwidth]{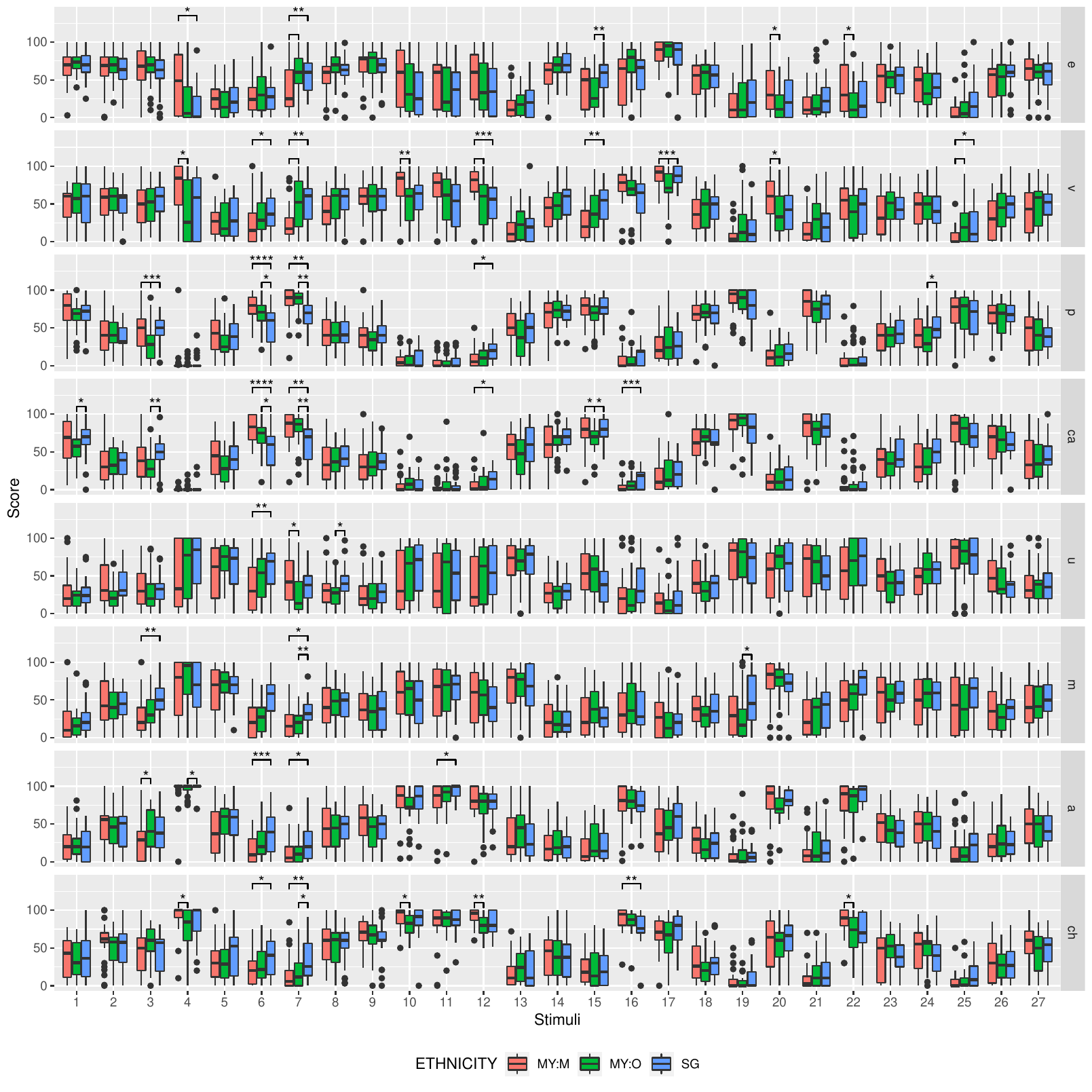}
    \caption{Summary plot of perceived affective quality scores across each stimuli. The braces indicate the significant pair via posthoc Conover-Iman test and the asterisks indicate the $p$-value significance levels (`****':0.01\%; `***':0.1\%; `**':1\%; `*':5\%;).}
    \label{fig:boxplot}
\end{figure*}


%% file: commands/appendix-start.tex
\setcounter{section}{0}
\renewcommand{\thesection}{Appendix \Alph{section}}

\setcounter{table}{0}
\renewcommand{\thetable}{\Alph{section}.\arabic{table}}

%% file: table/inittransPAQ.tex
\newcommand{\bigsq}{\scalebox{1.5}{$\square$}}

\begin{sffamily}
\setcounter{table}{1}
\small
\noindent\textbf{\color{scolor}Table \thetable}\par%
\noindent{\raggedright Translated instructions for the graphical user interface and perceived affective quality questionnaire (C.3.1.3, ISO~12913-2) in \ibm\ (ISO~639-3:~\sczsm). The upright text shows the \textsc{zsm} version as presented to the participants. The following italicised text in square brackets indicates the corresponding English text that was translated to \sczsm.}
\label{tab:inittransPAQ}
\small
\setlength{\tabcolsep}{0pt}
\vskip1em\noindent
\begin{tabularx}{\textwidth}{%
    l
    @{\hspace{-2.5em}}*{5}{>{\centering\arraybackslash}X}%
}
\toprule
    \multicolumn{6}{p{17cm}}{
        Tekan ``Mainkan bunyi'' untuk mendengar bunyi persekitaran semasa. Setelah selesai mendengar trek tersebut sekurang-kurangnya sekali, sila jawab soalan-soalan berikut. Anda perlu menekan setiap slider sekurang-kurangnya satu kali. Anda boleh mendengar trek seberapa kali yang diperlukan bagi menjawab soalan-soalan berikut. Harap maklum bahawa slider hanya akan muncul setelah trek berakhir. \newline
        [\textit{Press ``Play sound'' to hear the present sound environment. After listening to the track at least once, please answer the questions after you have finished listening to the track. You need to press all sliders at least 1 time. You may listen to the track for as many times as necessary to answer the questions. Please note that sliders will only appear after the track has ended.}]
    } \\
\midrule
    \multicolumn{6}{p{17cm}}{%
        Bagi setiap skala di bawah, sejauh manakah pendapat anda terhadap pengalaman bunyi persekitaran sebentar tadi adalah: \newline
        [\textit{For each scale below, to what extent do you think the sound environment you have just experienced is:}] 
    } \\
\midrule
    & Tidak sama sekali             
    &              
    &    
    &          
    & Amat sekali \\ 
    & [\textit{Strongly disagree}]                      
    &              
    &      
    &               
    & [\textit{Strongly agree}]   \\ 
\midrule
    Menyenangkan &
    0& & & &
    100\\
    {}[\textit{Stage 1 trans. ``pleasant''}]&
    \multicolumn{5}{c}{\SquareSLIDER{12cm}{0.5}} \\
    Huru-hara & 
    0& & & &
    100\\
    {}[\textit{Stage 1 trans. ``chaotic''}]& \multicolumn{5}{c}{\SquareSLIDER{12cm}{0.5}}\\
    Rancak & 
    0& & & &
    100\\
    {}[\textit{Stage 1 trans. ``vibrant''}]&
    \multicolumn{5}{c}{\SquareSLIDER{12cm}{0.5}} \\
    Tidak meriah & 
    0& & & &
    100\\
    {}[\textit{Stage 1 trans. ``uneventful''}]&
    \multicolumn{5}{c}{\SquareSLIDER{12cm}{0.5}}\\
    Tenang& 
    0& & & &
    100\\
    {}[\textit{Stage 1 trans. ``calm''}]&
    \multicolumn{5}{c}{\SquareSLIDER{12cm}{0.5}}\\
    Membingitkan; Menjengkelkan & 
    0& & & &
    100\\
    {}[\textit{Stage 1 trans. ``annoying''}]&
    \multicolumn{5}{c}{\SquareSLIDER{12cm}{0.5}}\\
    Meriah& 
    0& & & &
    100\\
    {}[\textit{Stage 1 trans. ``eventful''}]&
    \multicolumn{5}{c}{\SquareSLIDER{12cm}{0.5}}\\
    Membosankan& 
    0& & & &
    100\\
    {}[\textit{Stage 1 trans. ``monotonous''}] & \multicolumn{5}{c}{\SquareSLIDER{12cm}{0.5}}\\
\bottomrule
\end{tabularx}
\end{sffamily}

%% file: commands/next-appendix.tex
\setcounter{table}{1}
\renewcommand{\thetable}{\Alph{section}.\arabic{table}}
\renewcommand{\thefigure}{\Alph{section}.\arabic{figure}}

%% file: table/kwtTable.tex
\begin{table}[H]
\tiny
\caption{\label{tab:kwt}Summary of Kruskal-Wallis test results for ethnonational differences between individual PAQ attributes, \textit{ISOPL}, and \textit{ISOEV}. Asterisks indicate the $p$-value significance levels (`****':0.01\%; `***':0.1\%; `**':1\%; `*':5\%;) and parentheses indicate the effect size (`S':small; `M':medium; `L':large). }
\centering
\begin{tabularx}{\linewidth}{lXrrrrrrrrrr}
\toprule
Stimuli
& Statistic 
& \textit{e} 
& \textit{v} 
& \textit{p} 
& \textit{ca} 
& \textit{u} 
& \textit{m} 
& \textit{a} 
& \textit{ch} 
& \textit{ISOPL} 
& \textit{ISOEV}\\

\midrule
& pvalue & 0.8922 & 0.6866 & 0.0848 & *0.0229 & 0.8497 & 0.3356 & 0.8912 & 0.9657 & 0.5464 & 0.2345\\
\cmidrule{2-12}
\multirow[t]{-2}{*}{\raggedright\arraybackslash 1} & effect & (S)-0.0193 & (S)-0.0136 & (S)0.0319 & (M)0.0603 & (S)-0.0182 & 0.0020 & (S)-0.0192 & (S)-0.0210 & -0.0086 & 0.0098\\
\cmidrule{1-12}
 & pvalue & 0.7010 & 0.5245 & 0.6466 & 0.9061 & *0.0312 & 0.4859 & 0.5070 & 0.1692 & 0.1697 & 0.2938\\
\cmidrule{2-12}
\multirow[t]{-2}{*}{\raggedright\arraybackslash 2} & effect & (S)-0.0140 & -0.0077 & (S)-0.0123 & (S)-0.0196 & (S)0.0536 & -0.0060 & -0.0070 & (S)0.0169 & (S)0.0168 & 0.0049\\
\cmidrule{1-12}
 & pvalue & 0.7158 & 0.7277 & **0.0011 & **0.0044 & 0.6767 & **0.0048 & *0.0338 & 0.0966 & **0.0020 & 0.1467\\
\cmidrule{2-12}
\multirow[t]{-2}{*}{\raggedright\arraybackslash 3} & effect & (S)-0.0145 & (S)-0.0148 & (M)0.1256 & (M)0.0963 & (S)-0.0132 & (M)0.0941 & (S)0.0519 & (S)0.0291 & (M)0.1136 & (S)0.0200\\
\cmidrule{1-12}
 & pvalue & *0.0194 & *0.0169 & 0.8589 & 0.9519 & 0.3967 & 0.3482 & 0.1311 & *0.0240 & 0.4322 & *0.0322\\
\cmidrule{2-12}
\multirow[t]{-2}{*}{\raggedright\arraybackslash 4} & effect & (M)0.0640 & (M)0.0670 & (S)-0.0184 & (S)-0.0207 & -0.0016 & 0.0012 & (S)0.0224 & (S)0.0593 & -0.0035 & (S)0.0529\\
\cmidrule{1-12}
 & pvalue & 0.2984 & 0.5899 & 0.1018 & *0.0490 & 0.2225 & 0.5330 & 0.2140 & 0.0860 & *0.0269 & 0.3958\\
\cmidrule{2-12}
\multirow[t]{-2}{*}{\raggedright\arraybackslash 5} & effect & 0.0045 & (S)-0.0103 & (S)0.0279 & (S)0.0438 & (S)0.0109 & -0.0081 & (S)0.0118 & (S)0.0316 & (S)0.0568 & -0.0016\\
\cmidrule{1-12}
 & pvalue & 0.7964 & *0.0319 & ***0.0001 & ***0.0001 & **0.0065 & ****0.0000 & ***0.0004 & *0.0171 & ****0.0000 & 0.7393\\
\cmidrule{2-12}
\multirow[t]{-2}{*}{\raggedright\arraybackslash 6} & effect & (S)-0.0168 & (S)0.0531 & (L)0.1887 & (L)0.1909 & (M)0.0878 & (L)0.1996 & (L)0.1488 & (M)0.0667 & (L)0.2162 & (S)-0.0152\\
\cmidrule{1-12}
 & pvalue & **0.0033 & **0.0017 & ***0.0004 & ***0.0006 & *0.0226 & **0.0022 & *0.0193 & ***0.0009 & ***0.0006 & ***0.0003\\
\cmidrule{2-12}
\multirow[t]{-2}{*}{\raggedright\arraybackslash 7} & effect & (M)0.1023 & (M)0.1169 & (L)0.1464 & (L)0.1400 & (M)0.0606 & (M)0.1112 & (M)0.0640 & (M)0.1297 & (M)0.1386 & (L)0.1529\\
\cmidrule{1-12}
 & pvalue & 0.0795 & 0.0670 & 0.9389 & 0.1881 & *0.0213 & 0.9626 & 0.6044 & 0.7504 & 0.8042 & 0.2756\\
\cmidrule{2-12}
\multirow[t]{-2}{*}{\raggedright\arraybackslash 8} & effect & (S)0.0333 & (S)0.0370 & (S)-0.0204 & (S)0.0146 & (M)0.0619 & (S)-0.0209 & (S)-0.0108 & (S)-0.0155 & (S)-0.0170 & 0.0063\\
\cmidrule{1-12}
 & pvalue & 0.3322 & 0.8398 & 0.4943 & 0.4986 & 0.6499 & 0.5985 & 0.1561 & 0.3011 & 0.5723 & 0.7917\\
\cmidrule{2-12}
\multirow[t]{-2}{*}{\raggedright\arraybackslash 9} & effect & 0.0022 & (S)-0.0179 & -0.0064 & -0.0066 & (S)-0.0124 & (S)-0.0106 & (S)0.0186 & 0.0044 & -0.0096 & (S)-0.0167\\
\cmidrule{1-12}
 & pvalue & 0.1339 & **0.0099 & 0.9309 & 0.2726 & 0.1489 & 0.5283 & 0.0799 & *0.0176 & 0.7436 & 0.1012\\
\cmidrule{2-12}
\multirow[t]{-2}{*}{\raggedright\arraybackslash 10} & effect & (S)0.0220 & (M)0.0787 & (S)-0.0202 & 0.0065 & (S)0.0197 & -0.0079 & (S)0.0332 & (M)0.0661 & (S)-0.0153 & (S)0.0281\\
\cmidrule{1-12}
 & pvalue & 0.3501 & 0.0957 & 0.7691 & 0.5523 & 0.5499 & 0.8921 & *0.0225 & 0.6610 & 0.1256 & 0.1562\\
\cmidrule{2-12}
\multirow[t]{-2}{*}{\raggedright\arraybackslash 11} & effect & 0.0011 & (S)0.0293 & (S)-0.0160 & -0.0088 & -0.0087 & (S)-0.0193 & (M)0.0608 & (S)-0.0127 & (S)0.0234 & (S)0.0186\\
\cmidrule{1-12}
 & pvalue & 0.1490 & ***0.0002 & **0.0092 & *0.0473 & 0.2027 & 0.4829 & 0.2888 & **0.0047 & 0.5712 & **0.0091\\
\cmidrule{2-12}
\multirow[t]{-2}{*}{\raggedright\arraybackslash 12} & effect & (S)0.0196 & (L)0.1601 & (M)0.0801 & (S)0.0446 & (S)0.0130 & -0.0059 & 0.0053 & (M)0.0946 & -0.0096 & (M)0.0805\\
\cmidrule{1-12}
 & pvalue & 0.5336 & 0.4740 & 0.1234 & 0.1761 & 0.5949 & 0.9105 & 0.1603 & 0.3326 & 0.2151 & 0.3677\\
\cmidrule{2-12}
\multirow[t]{-2}{*}{\raggedright\arraybackslash 13} & effect & -0.0081 & -0.0055 & (S)0.0238 & (S)0.0160 & (S)-0.0105 & (S)-0.0197 & (S)0.0181 & 0.0022 & (S)0.0117 & 0.0000\\
\cmidrule{1-12}
 & pvalue & 0.1339 & *0.0394 & 0.7714 & 0.4814 & 0.5111 & 0.9514 & 0.8511 & 0.8423 & 0.3714 & 0.6545\\
\cmidrule{2-12}
\multirow[t]{-2}{*}{\raggedright\arraybackslash 14} & effect & (S)0.0220 & (S)0.0486 & (S)-0.0161 & -0.0058 & -0.0071 & (S)-0.0207 & (S)-0.0182 & (S)-0.0180 & -0.0002 & (S)-0.0125\\
\cmidrule{1-12}
 & pvalue & **0.0041 & **0.0079 & *0.0252 & **0.0093 & 0.1363 & 0.1101 & 0.3836 & 0.9885 & 0.0886 & *0.0185\\
\cmidrule{2-12}
\multirow[t]{-2}{*}{\raggedright\arraybackslash 15} & effect & (M)0.0976 & (M)0.0835 & (S)0.0582 & (M)0.0800 & (S)0.0216 & (S)0.0262 & -0.0009 & (S)-0.0215 & (S)0.0309 & (M)0.0649\\
\cmidrule{1-12}
 & pvalue & 0.0611 & 0.1087 & 0.1161 & **0.0017 & 0.1949 & 0.5456 & 0.8784 & *0.0153 & 0.4645 & 0.1364\\
\cmidrule{2-12}
\multirow[t]{-2}{*}{\raggedright\arraybackslash 16} & effect & (S)0.0390 & (S)0.0265 & (S)0.0251 & (M)0.1169 & (S)0.0138 & -0.0086 & (S)-0.0189 & (M)0.0691 & -0.0051 & (S)0.0216\\
\cmidrule{1-12}
 & pvalue & 0.1856 & **0.0048 & 0.9455 & 0.4601 & 0.2426 & 0.7042 & 0.1792 & 0.1366 & 0.7813 & 0.8305\\
\cmidrule{2-12}
\multirow[t]{-2}{*}{\raggedright\arraybackslash 17} & effect & (S)0.0149 & (M)0.0942 & (S)-0.0205 & -0.0049 & 0.0091 & (S)-0.0141 & (S)0.0156 & (S)0.0215 & (S)-0.0164 & (S)-0.0177\\
\cmidrule{1-12}
 & pvalue & 0.7283 & 0.3410 & 0.6358 & 0.3277 & 0.2039 & 0.4012 & 0.0962 & 0.2346 & *0.0403 & 0.7897\\
\cmidrule{2-12}
\multirow[t]{-2}{*}{\raggedright\arraybackslash 18} & effect & (S)-0.0148 & 0.0016 & (S)-0.0119 & 0.0025 & (S)0.0128 & -0.0019 & (S)0.0292 & 0.0098 & (S)0.0481 & (S)-0.0166\\
\cmidrule{1-12}
 & pvalue & 0.3957 & 0.1317 & 0.3188 & 0.1889 & 0.7906 & 0.0512 & 0.8993 & 0.4409 & 0.0797 & 0.7631\\
\cmidrule{2-12}
\multirow[t]{-2}{*}{\raggedright\arraybackslash 19} & effect & -0.0016 & (S)0.0223 & 0.0031 & (S)0.0145 & (S)-0.0166 & (S)0.0429 & (S)-0.0194 & -0.0039 & (S)0.0332 & (S)-0.0159\\
\cmidrule{1-12}
 & pvalue & *0.0320 & *0.0188 & 0.5063 & 0.7328 & 0.1162 & 0.4963 & *0.0429 & 0.4304 & 0.7258 & *0.0302\\
\cmidrule{2-12}
\multirow[t]{-2}{*}{\raggedright\arraybackslash 20} & effect & (S)0.0531 & (M)0.0647 & -0.0069 & (S)-0.0150 & (S)0.0251 & -0.0065 & (S)0.0467 & -0.0034 & (S)-0.0148 & (S)0.0544\\
\cmidrule{1-12}
 & pvalue & 0.4372 & 0.2387 & 0.1229 & 0.2525 & 0.3195 & 0.2694 & 0.6662 & 0.1848 & 0.5034 & 0.3455\\
\cmidrule{2-12}
\multirow[t]{-2}{*}{\raggedright\arraybackslash 21} & effect & -0.0038 & 0.0094 & (S)0.0238 & 0.0082 & 0.0031 & 0.0068 & (S)-0.0129 & (S)0.0150 & -0.0068 & 0.0014\\
\cmidrule{1-12}
 & pvalue & *0.0283 & 0.3071 & 0.7082 & 0.4818 & 0.2880 & *0.0151 & 0.2179 & *0.0123 & 0.4669 & **0.0051\\
\cmidrule{2-12}
\multirow[t]{-2}{*}{\raggedright\arraybackslash 22} & effect & (S)0.0557 & 0.0039 & (S)-0.0142 & -0.0059 & 0.0053 & (M)0.0694 & (S)0.0114 & (M)0.0738 & -0.0052 & (M)0.0929\\
\cmidrule{1-12}
 & pvalue & 0.9169 & 0.2876 & 0.6474 & 0.1744 & 0.3075 & 0.5249 & 0.1535 & 0.2067 & 0.3664 & 0.1745\\
\cmidrule{2-12}
\multirow[t]{-2}{*}{\raggedright\arraybackslash 23} & effect & (S)-0.0199 & 0.0054 & (S)-0.0123 & (S)0.0162 & 0.0039 & -0.0077 & (S)0.0190 & (S)0.0125 & 0.0001 & (S)0.0162\\
\cmidrule{1-12}
 & pvalue & 0.3843 & 0.6265 & *0.0331 & *0.0305 & 0.2755 & 0.6393 & 0.4755 & 0.0738 & 0.1511 & 0.1465\\
\cmidrule{2-12}
\multirow[t]{-2}{*}{\raggedright\arraybackslash 24} & effect & -0.0009 & (S)-0.0116 & (S)0.0523 & (S)0.0541 & 0.0063 & (S)-0.0120 & -0.0056 & (S)0.0349 & (S)0.0193 & (S)0.0200\\
\cmidrule{1-12}
 & pvalue & 0.0529 & **0.0052 & 0.6339 & 0.2347 & 0.9759 & 0.0857 & 0.1710 & 0.0545 & 0.1953 & 0.3919\\
\cmidrule{2-12}
\multirow[t]{-2}{*}{\raggedright\arraybackslash 25} & effect & (S)0.0421 & (M)0.0925 & (S)-0.0118 & 0.0098 & (S)-0.0212 & (S)0.0317 & (S)0.0167 & (S)0.0415 & (S)0.0138 & -0.0014\\
\cmidrule{1-12}
 & pvalue & 0.0960 & 0.1459 & 0.9533 & 0.8900 & 0.1443 & 0.3838 & 0.9719 & 0.9136 & 0.7185 & 0.0536\\
\cmidrule{2-12}
\multirow[t]{-2}{*}{\raggedright\arraybackslash 26} & effect & (S)0.0292 & (S)0.0201 & (S)-0.0207 & (S)-0.0192 & (S)0.0203 & -0.0009 & (S)-0.0211 & (S)-0.0198 & (S)-0.0146 & (S)0.0419\\
\cmidrule{1-12}
 & pvalue & 0.8821 & 0.8291 & 0.7761 & 0.3570 & 0.9775 & 0.4754 & 0.6613 & 0.2503 & 0.9078 & 0.7823\\
\cmidrule{2-12}
\multirow[t]{-2}{*}{\raggedright\arraybackslash 27} & effect & (S)-0.0190 & (S)-0.0177 & (S)-0.0162 & 0.0006 & (S)-0.0212 & -0.0056 & (S)-0.0127 & 0.0084 & (S)-0.0196 & (S)-0.0164\\
\bottomrule

\end{tabularx}
\end{table}

%% file: table/citTable.tex
\small
\begin{longtable}{p{1.2cm}p{8cm}*{3}{>{\raggedleft\arraybackslash}p{2cm}}}

\caption{Summary of posthoc Conover-Iman test results for ethnonational differences between individual PAQ attributes, \iisopl, and \iisoev. Asterisks indicate the $p$-value significance levels (`****':0.01\%; `***':0.1\%; `**':1\%; `*':5\%;)}
\label{tab:cit}
\endfirsthead
\toprule
Stimuli & PAQ & MY:M--MY:O & MY:M--SG & MY:O--SG\\
\midrule
\endhead
    \\
    \multicolumn{5}{r}{[Continued on next page]} \\
\endfoot
\endlastfoot

\toprule
Stimuli & PAQ & MY:M--MY:O & MY:M--SG & MY:O--SG\\
\midrule
1 & calm & 0.1164 & 1.0000 & *0.0250\\
\cmidrule{1-5}
2 & uneventful & 0.0598 & 1.0000 & 0.0689\\
\cmidrule{1-5}
 & pleasant & *0.0372 & 0.6028 & ***0.0006\\
\cmidrule{2-5}
 & calm & 0.5180 & 0.1319 & **0.0026\\
\cmidrule{2-5}
 & monotonous & 0.8094 & **0.0034 & 0.0771\\
\cmidrule{2-5}
 & annoying & *0.0272 & 0.3972 & 0.7508\\
\cmidrule{2-5}
\multirow[t]{-5}{*}{\raggedright\arraybackslash 3} & \iisopl & **0.0022 & 1.0000 & *0.0132\\
\cmidrule{1-5}
 & eventful & 0.0604 & *0.0284 & 1.0000\\
\cmidrule{2-5}
 & vibrant & *0.0169 & 0.0981 & 1.0000\\
\cmidrule{2-5}
 & chaotic & *0.0179 & 0.4910 & 0.4746\\
\cmidrule{2-5}
\multirow[t]{-4}{*}{\raggedright\arraybackslash 4} & \iisoev & *0.0348 & 0.1494 & 1.0000\\
\cmidrule{1-5} 
 & calm & 0.0596 & 1.0000 & 0.1829\\
\cmidrule{2-5} 
\multirow[t]{-2}{*}{\raggedright\arraybackslash 5} & \iisopl & *0.0205 & 0.4474 & 0.5667\\
\cmidrule{1-5}
 & vibrant & 0.3542 & *0.0258 & 0.8010\\
\cmidrule{2-5}
 & pleasant & 0.1657 & ****0.0000 & *0.0129\\
\cmidrule{2-5}
 & calm & 0.1204 & ****0.0000 & *0.0169\\
\cmidrule{2-5}
 & uneventful & 0.2153 & **0.0039 & 0.4031\\
\cmidrule{2-5}
 & monotonous & 1.0000 & ***0.0001 & ***0.0002\\
\cmidrule{2-5}
 & annoying & 0.0535 & ***0.0001 & 0.1870\\
\cmidrule{2-5}
 & chaotic & 0.3694 & *0.0120 & 0.4876\\
\cmidrule{2-5}
\multirow[t]{-8}{*}{\raggedright\arraybackslash 6} & \iisopl & 0.4072 & ****0.0000 & **0.0013\\
\cmidrule{1-5}
 & eventful & **0.0088 & **0.0069 & 1.0000\\
\cmidrule{2-5}
 & vibrant & **0.0050 & **0.0035 & 1.0000\\
\cmidrule{2-5}
 & pleasant & 1.0000 & **0.0020 & ***0.0007\\
\cmidrule{2-5}
 & calm & 1.0000 & **0.0014 & **0.0017\\
\cmidrule{2-5}
 & uneventful & *0.0234 & 1.0000 & 0.1275\\
\cmidrule{2-5}
 & monotonous & 1.0000 & *0.0102 & **0.0032\\
\cmidrule{2-5}
 & annoying & 1.0000 & *0.0310 & 0.0567\\
\cmidrule{2-5}
 & chaotic & 0.8092 & ***0.0006 & *0.0194\\
\cmidrule{2-5}
 & \iisopl & 0.9674 & *0.0119 & ***0.0004\\
\cmidrule{2-5}
\multirow[t]{-10}{*}{\raggedright\arraybackslash 7} & \iisoev & ***0.0009 & ***0.0008 & 1.0000\\
\cmidrule{1-5}
8 & uneventful & 0.4235 & 0.5355 & *0.0156\\
\cmidrule{1-5}
 & vibrant & **0.0080 & 0.0969 & 1.0000\\
\cmidrule{2-5}
\multirow[t]{-2}{*}{\raggedright\arraybackslash 10} & chaotic & *0.0140 & 0.8847 & 0.1973\\
\cmidrule{1-5} \pagebreak
11 & annoying & 0.6785 & *0.0171 & 0.3221\\
\cmidrule{1-5}
 & vibrant & ***0.0009 & ***0.0004 & 1.0000\\
\cmidrule{2-5}
 & pleasant & 1.0000 & *0.0104 & 0.0509\\
\cmidrule{2-5}
 & calm & 1.0000 & *0.0497 & 0.2341\\
\cmidrule{2-5}
 & chaotic & **0.0030 & 0.1039 & 0.6220\\
\cmidrule{2-5}
\multirow[t]{-5}{*}{\raggedright\arraybackslash 12} & \iisoev & *0.0273 & *0.0149 & 1.0000\\
\cmidrule{1-5}
14 & vibrant & 0.6050 & *0.0320 & 0.5586\\
\cmidrule{1-5}
 & eventful & 0.7860 & 0.0767 & **0.0028\\
\cmidrule{2-5}
 & vibrant & 0.1641 & **0.0051 & 0.5959\\
\cmidrule{2-5}
 & pleasant & 0.0831 & 1.0000 & *0.0358\\
\cmidrule{2-5}
 & calm & *0.0380 & 1.0000 & *0.0127\\
\cmidrule{2-5}
\multirow[t]{-5}{*}{\raggedright\arraybackslash 15} & \iisoev & 1.0000 & *0.0304 & 0.0535\\
\cmidrule{1-5}
 & calm & 0.2381 & ***0.0008 & 0.1330\\
\cmidrule{2-5}
\multirow[t]{-2}{*}{\raggedright\arraybackslash 16} & chaotic & 0.8310 & *0.0118 & 0.1896\\
\cmidrule{1-5}
17 & vibrant & **0.0036 & 0.9191 & 0.0656\\
\cmidrule{1-5}
18 & \iisopl & *0.0374 & 1.0000 & 0.2783\\
\cmidrule{1-5}
 & eventful & *0.0254 & 0.3972 & 0.7203\\
\cmidrule{2-5}
 & vibrant & *0.0213 & 0.0856 & 1.0000\\
\cmidrule{2-5}
 & annoying & 0.0548 & 1.0000 & 0.1507\\
\cmidrule{2-5}
\multirow[t]{-4}{*}{\raggedright\arraybackslash 20} & \iisoev & *0.0246 & 0.8266 & 0.3273\\
\cmidrule{1-5}
 & eventful & *0.0340 & 0.1110 & 1.0000\\
\cmidrule{2-5}
 & monotonous & 1.0000 & *0.0263 & *0.0418\\
\cmidrule{2-5}
 & chaotic & *0.0179 & *0.0414 & 1.0000\\
\cmidrule{2-5}
\multirow[t]{-4}{*}{\raggedright\arraybackslash 22} & \iisoev & *0.0190 & **0.0076 & 1.0000\\
\cmidrule{1-5}
 & pleasant & 0.2611 & 1.0000 & *0.0295\\
\cmidrule{2-5}
\multirow[t]{-2}{*}{\raggedright\arraybackslash 24} & calm & 1.0000 & 0.1096 & *0.0399\\
\cmidrule{1-5}
25 & vibrant & *0.0110 & *0.0129 & 1.0000\\
\bottomrule
\end{longtable}